\documentclass[reprint,floatfix]{revtex4-2}
\usepackage[utf8]{inputenc}
\usepackage{amsmath}
\usepackage{amsfonts}
\usepackage{amssymb}
\usepackage{graphicx}
\usepackage[T1]{fontenc}
\usepackage{braket}
\usepackage{hyperref}
\usepackage{natbib}
\begin{document}

\preprint{APS/123-QED}

\title{Quantum Repeater using Two-Mode Squeezed States and Atomic Noiseless Amplifiers}

\author{Anders J. E. Bjerrum}
    \affiliation{Center for Macroscopic Quantum States (bigQ), Department of Physics, Technical University of Denmark, 2800 Kongens Lyngby, Denmark}
\author{Jonatan B. Brask}%
    \affiliation{Center for Macroscopic Quantum States (bigQ), Department of Physics, Technical University of Denmark, 2800 Kongens Lyngby, Denmark}
\author{Jonas S. Neergaard-Nielsen}%
    \affiliation{Center for Macroscopic Quantum States (bigQ), Department of Physics, Technical University of Denmark, 2800 Kongens Lyngby, Denmark}
\author{Ulrik L. Andersen}%
    \affiliation{Center for Macroscopic Quantum States (bigQ), Department of Physics, Technical University of Denmark, 2800 Kongens Lyngby, Denmark}

\date{\today}

\begin{abstract}

We perform a theoretical investigation into how a two-mode squeezed vacuum state, that has undergone photon loss, can be stored and purified using noiseless amplification with a collection of solid-state qubits. The proposed method may be used to probabilistically increase the entanglement between the two parties sharing the state. The proposed amplification step is similar in structure to a set of quantum scissors. However, in this work the amplification step is realized by a state transfer from an optical mode to a set of solid-state qubits, which act as a quantum memory. We explore two different applications, the generation of entangled many-qubit registers, and the construction of quantum repeaters for long-distance quantum key distribution.

\end{abstract}

\maketitle

\section{Introduction}
Quantum communication is the act of distributing quantum states in a network \cite{gisin7}. It enables the generation of secret encryption keys \cite{pirandola20}, and perhaps, the establishment of a fully fault-tolerant quantum internet  \cite{kimble8,wehner18}. The different nodes of the network are usually connected by photonic communication channels owing to the weak influence of the environment on the coherence of optical photons. However, photons suffer from propagation loss with the probability of successful transmission decaying exponentially with distance.\\

\begin{figure}
\centering
\includegraphics[width=0.9\columnwidth]{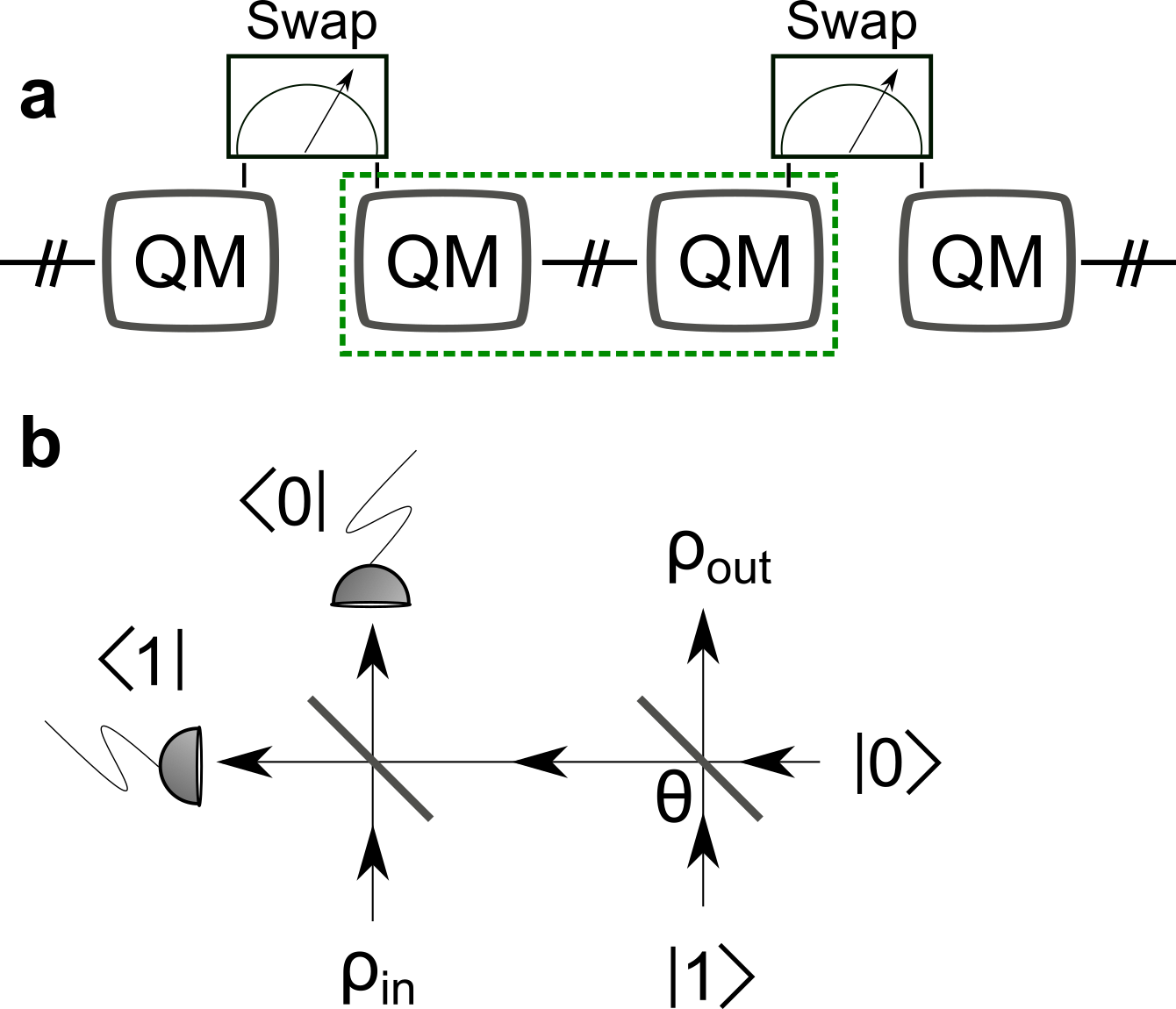}
\caption{\textit{\textbf{a}: Repeater scheme with entanglement swaps (meters) acting on sets of quantum memories (QM) separated by lossy channels. A single repeater segment is enclosed by a green box, and will include a purification step. \textbf{b}: Layout of a quantum scissor. The left beamsplitter is balanced (50:50) and the right beamsplitter is tunable with transmission $\cos(\theta)^2$. The transmission may be tuned to purify the state $\rho_{in}$ at the single photon level.}}
\label{Intro:Figure:repScissor}
\end{figure}

The exponential scaling can be mitigated using quantum repeater nodes between the sender and receiver stations leading to polynomial ~\cite{briegel98,duan01} or even constant-rate scaling \cite{jiang16} for schemes based on error correction. We consider a quantum repeater architechture with two-way classical communication and without error correction, as originally envisioned ~\cite{briegel98,duan01}.
In this scheme, entanglement between sender and receiver is established by first distributing and purifying entangled states over shorter segments. These entangled segments then undergo a series of entanglement swaps, ultimately generating entanglement between sender and receiver (see Fig. \ref{Intro:Figure:repScissor}a). However, due to the probabilistic nature of the purification protocol, quantum memories must be placed at each repeater node to attain the polynomial scaling. Many different platforms have been considered for memories in quantum repeaters including atomic ensembles \cite{sangouard11}, trapped ions \cite{sangouard09}, solid-state systems \cite{usmani10}, and mechanical resonators \cite{wallucks20}. The basic structure of all quantum repeaters is largely independent of the type of memory employed, however.\\

One intriguing approach for the probabilistic purification of a quantum state is the protocol of noiseless linear amplification \cite{ralph09}. It has mainly been applied in continuous-variable (CV) quantum repeater schemes to enable long-distance distribution of quadrature and photon-number entanglement \cite{ralph11,dias17,dias18,seshadreesan19,seshadreesan20,ghalaii20,dias21} (see also \cite{furrer18} for a different approach). In its simplest form, the noiseless linear amplifier consists of a single quantum scissor scheme~\cite{barnett98} illustrated in Fig. \ref{Intro:Figure:repScissor}b. A single photon is split on a beam splitter to form an entangled state which is subsequently used to purify an input state ($\rho_{\texttt{in}}$) via quantum teleportation in a truncated two-dimensional Hilbert space. The achieved purification can be understood intuitively through the fact, that the projective measurement of the photon-detectors lower the entropy of the system, while the effectuated quantum teleportation preserve the coherences of the input state. By combining quantum scissor operations with quantum memories and entanglement swapping via Bell measurements, a quantum repeater network with polynomial loss scaling can be established. In previous CV quantum repeater protocols, the quantum scissors, the quantum memories and the Bell measurements are typically considered as being individual and independent physical elements. \\
In the present work we show that by using light-matter entangled states (for example generated by Nitrogen-Vacancy centers in diamond \cite{rozpedek19,maurer12,doherty13}), it is possible to perform noiseless linear amplification and storage of the quantum state in a single operation. In section \ref{Analysis:Section:SingleRep} we introduce the operation of noiseless linear amplification based on a photo-active qubit. We first consider the case with a single qubit, similar to a single quantum scissor operation, and subsequently generalize it to multiple qubits to explore the effect of noiseless linear amplification in a larger Hilbert space. In section \ref{Analysis:Section:Entanglement} we investigate the entanglement generated by our protocol and measure it using the negativity. In section \ref{Analysis:section:entSwap} we then introduce the structure of the quantum repeater scheme, including entanglement swapping. In section \ref{Analysis:section:results} we present the results in terms of secret key rates and Bell inequality violations.

\section{Analysis of a single repeater segment}
\label{Analysis:Section:SingleRep}

\begin{figure}
\centering
\includegraphics[width=1\columnwidth]{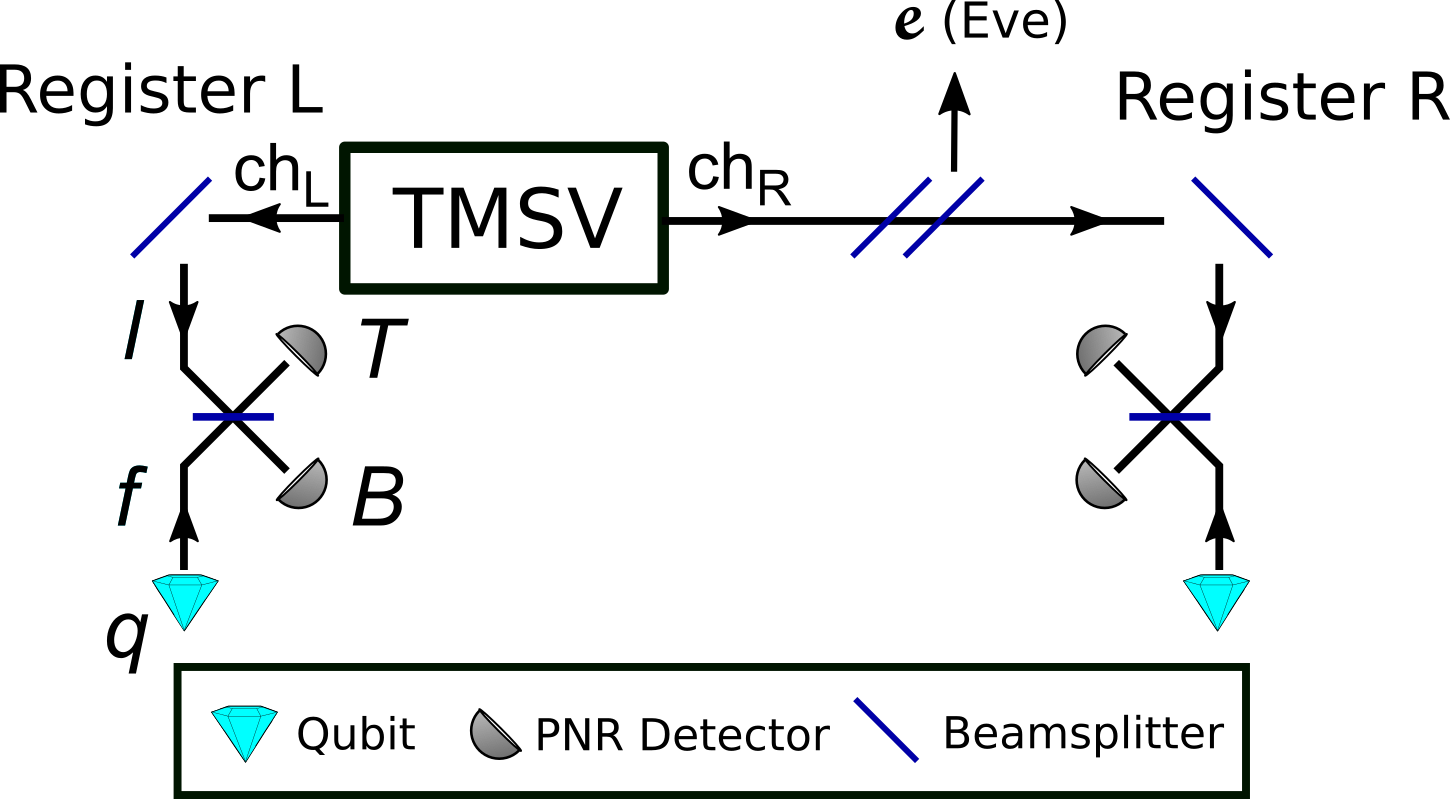}
\caption{\textit{Layout of the entanglement-sharing scheme, with a single qubit in each register (quantum memory). $\mathbf{e}$ is an environmental mode that couples to the fiber. The drawn setup corresponds to a repeater segment (the green dotted box in Fig. \ref{Intro:Figure:repScissor})}}
\label{Analysis:Figure:LayoutSingle}
\end{figure}

We start by presenting the repeater segment that forms the core of the quantum repeater protocol. It corresponds to the part of the repeater array enclosed by a green dotted box in Fig. \ref{Intro:Figure:repScissor}a and is schematically shown in Fig. \ref{Analysis:Figure:LayoutSingle} with a single solid state qubit (diamond) in each register. Our repeater scheme is based on the distribution of two-mode squeezed vacuum states followed by noiseless linear amplification and memorization by means of a photo-active three-level atomic systems. The basic idea is that the atomic systems produce spin-photon entanglement to be used as the resource for heralded noiseless amplification similar to the all-optical approach in Fig. \ref{Intro:Figure:repScissor}b where single-photon entanglement is used as the resource. However, in contrast to the pure optical approach in Fig. \ref{Intro:Figure:repScissor}b where the state is teleported onto another optical mode, in our scheme the state is teleported (and truncated) into a spin degree of freedom of the atomic system, and thus directly memorized after purification. While the atomic system could be realised by many different physical systems, here we focus on the Nitrogen-Vacancy center (NV). In this case the information is stored in the electronic spin degree of freedom of the NV center but it can also be swapped to a nearby (and very long-lived) $^{13}$C nuclear spin \cite{maurer12}. In addition to extending the lifetime, the swap also free up the electronic spin for a subsequent entangling round and it allows for entanglement swapping to be carried through Bell measurements between the electronic and nuclear spins (as discussed in section \ref{Analysis:section:entSwap}).\\

A two-mode squeezed vacuum (TMSV) state shares quadrature and photon number correlations between the left (L) and right (R) registers. The TMSV state is expressed in the photon-number basis as
\begin{equation}
    \ket{\texttt{TMSV}} = \sum_{n=0}^\infty c_n \ket{n,n},
\end{equation}
where the amplitudes $c_n$ are the Fock-state amplitudes
\begin{equation}
	c_n = (-e^{i \phi})^n  \sqrt{\frac{\langle n \rangle^n}{(1+\langle n \rangle )^{n+1}}} ,
\end{equation}
$\langle n \rangle$ is average photon number in each of the two modes and $\phi$ is the phase.

The register qubits have a dark state ($\ket{0}_q$) and a bright state ($\ket{1}_q$). The bright state emits a single photon into the optical mode $f$ when excited by some external mechanism, such as a driving laser, whereas the dark state never emits a photon. We initialize the qubit $q$ and optical mode $f$ in the state
\begin{align}
	\ket{q,f} = \cos(\theta)\ket{0}_q \ket{0}_f + \sin(\theta)\ket{1}_q \ket{0}_f ,
	\label{thetaIntro}
\end{align} 
with $\ket{0}_f$ being the optical vacuum state. Then we assume we can excite the qubit such that it emits a photon if it is in the bright state, thereby preparing the state
\begin{align}
	\cos(\theta)\ket{0}_q \ket{0}_f + \sin(\theta)\ket{1}_q \ket{1}_f .
\end{align}
States such as this one were produced experimentally using a NV center in ref. \cite{bernien13}.
We assume that the photons of the TMSV field are indistinguishable from the photons emitted by the qubits. Realistically, this may be a challenge to achieve, but can in principle be done with proper light source engineering and filtering. At each register the TMSV field is mixed with the field emitted by the register qubit on a balanced beamsplitter. Two photon-number-resolving detectors (PNR detectors) measure the outputs of the beamsplitter, and events where exactly one photon is detected at each register are considered successful. If we assume no loss in channels ch$_L$ and ch$_R$, then the qubit registers are projected into the entangled state
\begin{align}
    \ket{\psi} = \frac{1}{2} c_1 \cos(\theta_L) \cos(\theta_R) \ket{00} + \nonumber \\ 
    \frac{1}{2} c_0 \sin(\theta_L) \sin(\theta_R) \ket{11}  ,
\end{align}
where $\theta_{L}$ and $\theta_{R}$ are the superposition angles given in Eq. \ref{thetaIntro}, for the left and right qubit respectively. We note that $4\braket{\psi|\psi}$ will correspond to the probability that the projective measurements (photon detection) in register L and R succeed. The factor of 4 originates from the fact that the projective measurement can succeed in 2 different ways at both registers, i.e. either the top (T) or bottom (B) detector can register a single photon. We note that whether the bottom or top detector clicks, will influence the phase of the quantum state, and we assume that this is corrected for. \\

We then set the transmission of channel ch$_\texttt{R}$, connecting the TMSV source and register R, to $\eta_R$. We will for now assume that the channel ch$_\texttt{L}$ is lossless. We find that under these conditions, the density matrix describing register L and R, is given by
\begin{widetext}
\begin{align}
    \rho =  \frac{\mathcal{K}_1}{4} &\begin{pmatrix}
    |c_1|^2 \eta_R c_L^{2}  c_R^{2} & 0 & 0 &  c_1 c_0^* \sqrt{\eta_R} s_R s_L c_R c_L \\
    0 & |c_1|^2 (1-\eta_R) c_L^2 s_R^2 & 0 & 0 \\
    0 & 0 & 0 & 0 \\
    c_1^* c_0 \sqrt{\eta_R} s_R s_L c_R c_L & 0 & 0 & |c_0|^2 s_L^2 s_R^2 \\
    \end{pmatrix} ,
    \label{Analysis:Equation:1QubitLossy}
\end{align}
\end{widetext}
Where $s_L = \sin(\theta_L)$, $s_R = \sin(\theta_R)$, $c_L = \cos(\theta_L)$, and $c_R = \cos(\theta_R)$.
A derivation of this result can be found in Appendix A. The basis vectors describing the state are $ \ket{0}_L \ket{0}_R$, $\ket{0}_L \ket{1}_R$, $\ket{1}_L \ket{0}_R$, $\ket{1}_L \ket{1}_R $. Eg. the matrix element $\rho_{22} = \frac{\mathcal{K}_1}{4} |c_1|^2 (1-\eta_R) \cos(\theta_L)^2 \sin(\theta_R)^2$ corresponds to the state $\ket{0}_L \ket{1}_R \bra{0}_L \bra{1}_R$. $\mathcal{K}_1$ is the normalization and $4/\mathcal{K}_1$ will be the probability that the projective measurements in register L and R succeed. \\The diagonal term $\rho_{22}$ describes the situation where a photon is lost in the right channel. Suppose that the TMSV source emits a single photon into both ch$_L$ and ch$_R$, and that the photon is lost from channel ch$_R$. Then a successful measurement at register R and L implies that the qubit in register L is in the dark state ($\ket{0}_L$), and that the qubit in register R is in the bright state ($\ket{1}_R$).\\
In section \ref{Analysis:Section:Entanglement} we show that tuning of the angles $\theta_{L}$ and $\theta_{R}$, can increase (or decrease) the entanglement shared between the two registers, in a similar fashion as a pair of quantum scissors would. \\

\begin{figure*}
\centering
\includegraphics[width=2\columnwidth]{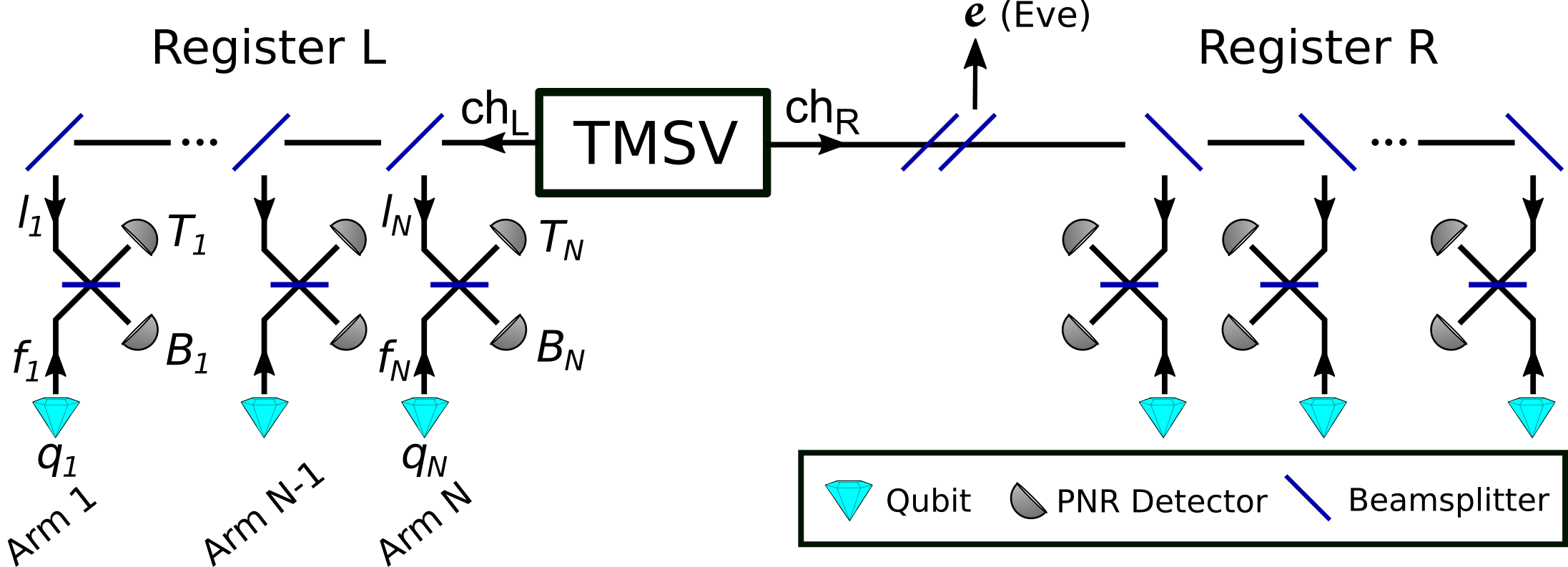}
\caption{\textit{Layout of the entanglement-sharing scheme investigated in this work. Entanglement is distributed to the atomic qubits consituting register $R$ and $L$ via a two-mode squeezed vacuum state.}}
\label{Analysis:Figure:Layout}
\end{figure*}

In the architecture discussed above and illustrated in Fig. \ref{Analysis:Figure:LayoutSingle}, the amount of distributed and purified entanglement is limited due to the restricted four-dimensional Hilbert space spanned by the two qubits. To circumvent this limitation we consider an expanded version of the two registers where every register now comprises several qubits and thus enlarges the dimensionality of the quantum memory. The setup can be seen in Fig. \ref{Analysis:Figure:Layout}. At the left register the TMSV state is split evenly into the $N$ arms of the register. Concurrently with this splitting of the TMSV, we excite the qubits, such that they will emit a photon if they are in the bright state. Again, PNR detectors measure on the output, and events where exactly one photon is detected in each of the N register arms, are considered successful. Conditioned on all the measurements succeeding, we obtain correlations between the number of bright-state qubits in the left register, and the number of photons in the right part of the TMSV state. Repeating the procedure at the right register ultimately creates entanglement between the two registers.\\

Our analysis, given in appendix A, reveals that the quantum state of the two many-qubit registers, when assuming no loss in the connecting channel, is given by
\begin{align}
	\ket{\alpha} = \mathcal{N}_N \sum_{n=0}^{N} c_n  \Delta(n,\theta_L) \Delta(n,\theta_R)  \ket{\mathbf{I}_{N-n}}_L \ket{\mathbf{I}_{N-n}}_R
\end{align}
The subscript $R/L$ indicates whether we are referring to the qubits in the left (L) or right (R) register. The superposition angle $\theta_{R/L}$ is assumed to be the same for all qubits in the same register. The vectors $\ket{\mathbf{I}_{N-n}}$ are even superpositions of all states containing $N-n$ bright state qubits and $n$ dark state qubits:
\begin{align}
    \ket{\mathbf{I}_{N-n}} = \binom{N}{n}^{-1/2}\sum_{\mathbf{i}_{N-n}} \ket{ \mathbf{i}_{N-n}}
\end{align}
where the sum runs over binary lists $\mathbf{i}_{N-n}$ of length $N$ with $N-n$ ones. $\mathcal{N}_N$ is the normalization, given by
\begin{align}
    \mathcal{N}_N^{-2} = \sum_{m=0}^{N} |c_m|^2 |\Delta(m,\theta_L)|^2 |\Delta(m,\theta_R)|^2 
\end{align}
The value of $\mathcal{N}_N^{-2}$ reflects the probability that the entanglement sharing scheme succeeds. The amplitude $\Delta(n,\theta)$ is
\begin{align}
	\Delta(n,\theta) = \sqrt{ \frac{N!} {2^N N^n (N-n)!} } \beta(n,\theta),
\end{align}
where $\beta(n,\theta)$ is
\begin{equation}
    \beta(n,\theta) = \cos(\theta)^{n} \sin(\theta)^{N-n}
\end{equation}
When including loss in both channels, we find the density matrix describing the two registers to be:
\begin{align}
\label{NoError:Equation:TracedDensityLoss_Analysis}
    \rho = & \sum_{n,m=0}^{\infty} \sum_{l,r=0}^{\texttt{min}(n,m)}  \Lambda(n,m,l,r) \cdot \\ &\ket{\mathbf{I}_{N-n+l}}_L \bra{\mathbf{I}_{N-m+l}} \otimes   \ket{\mathbf{I}_{N-n+r}}_R \bra{\mathbf{I}_{N-m+r}} \nonumber
\end{align}
This density matrix is not normalized, and the norm should be interpreted as the probability that entanglement sharing succeeds. The matrix elements are given by
\begin{widetext}
\begin{align}
    \Lambda(n,m,l,r) &= c_n c_m^* \epsilon_R(n,r) \epsilon_L(n,l) \epsilon_R(m,r)^* \epsilon_L(m,l)^*  \nonumber \\& \Delta(n-l,\theta_L) \Delta(n-r,\theta_R) \Delta(m-l,\theta_L)^* \Delta(m-r,\theta_R)^*
    \nonumber \\& \Theta(N+l-n)\Theta(N+r-n)\Theta(N+l-m)\Theta(N+r-m) ,
\end{align}
\end{widetext}
where $\Theta(x)$ is the step function,
\begin{align}
    \Theta(x) =     \begin{cases}
        1 & \text{if } x \geq 0\\
        0 & \text{if } x < 0
    \end{cases}
\end{align}
$\epsilon(n,l)$ is related to the transmission of the channel, $\eta_{R/L}$ (for ch$_\texttt{L}$ and ch$_\texttt{R}$), through the relation,
\begin{equation}
	\epsilon_{R/L}(n,l) = \sqrt{\binom{n}{n-l}} \eta_{R/L}^{(n-l)/2} (1-\eta_{R/L})^{l/2}
\end{equation}

\section{Entanglement of the Registers}
\label{Analysis:Section:Entanglement}
Based on the above analysis, we now evaluate the amount of entanglement between the two registers using negativity as the measure of entanglement. The negativity is defined as the absolute value of the sum of negative eigenvalues of the partial transpose of the density matrix, and can be shown to be an entanglement monotone \cite{vidal02}. If we have the density matrix $\rho$, then the partial transpose with respect to Alice's subsystem, $\rho^{T_A}$, has the matrix elements $\braket{i_A,j_B|\rho^{T_A}|k_A,l_B} = \braket{k_A,j_B|\rho|i_A,l_B}$. Given that $\rho^{T_A}$ has the negative eigenvalues $\mu_i$, then the negativity of the state $\rho$ is defined as
\begin{align}
    \texttt{Neg}(\rho) = |\sum_i \mu_i|
\end{align}

\begin{figure}
\centering
\includegraphics[width=1\columnwidth]{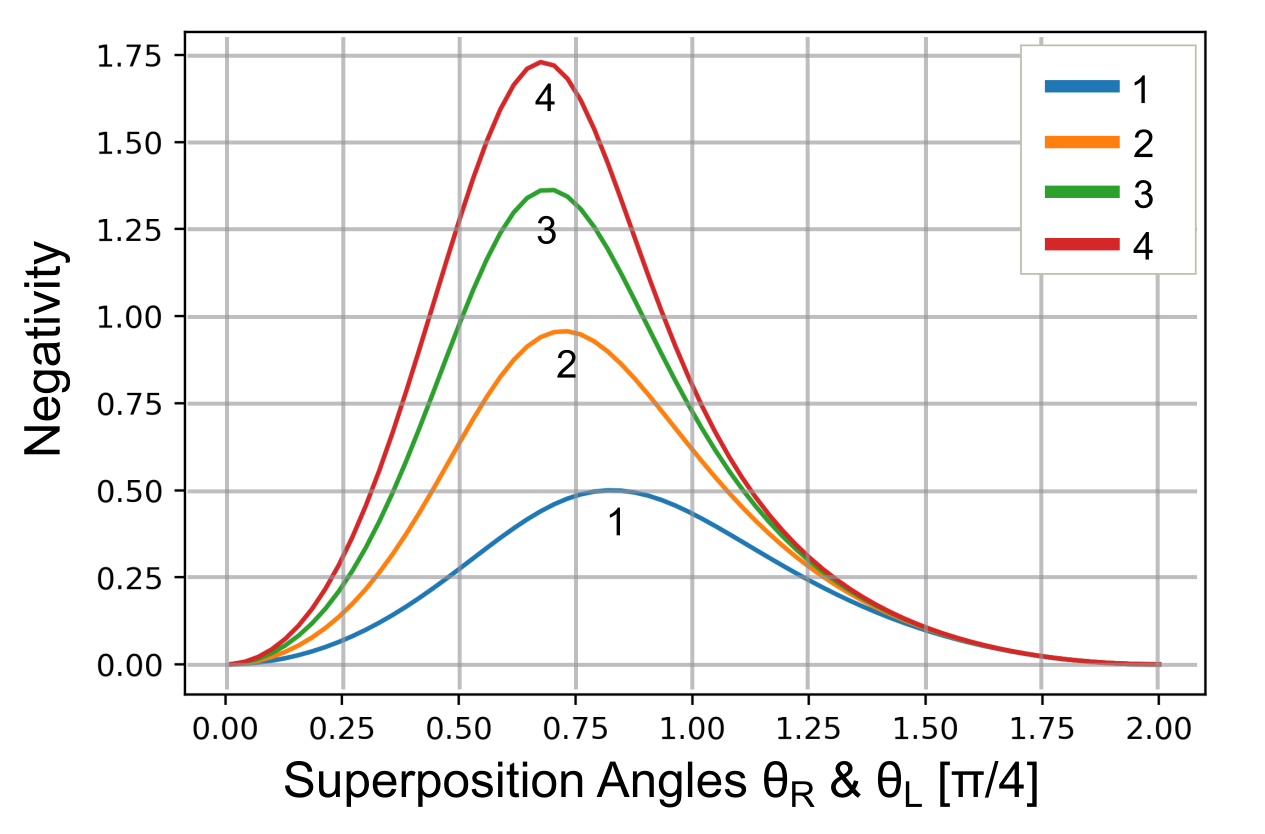}
\caption{\textit{The negativity of the two registers as a function of the superposition angle $\theta=\theta_R=\theta_L$. The different plots correspond to different number of qubits in the registers. The average number of photons $\langle n \rangle$ emitted by the TMSV source into each channel is fixed at 0.5. The superposition angle $\theta$ is shown along the x-axis in units of $\pi/4$.}}
\label{Results:Figure:NegVsTheta}
\end{figure}

The negativity is the same regardless of which party was transposed, since $(\rho^{T_A})^T = \rho^{T_B}$ and $\rho^{T_A}$ is hermitian. We will first assume the channels to be loss-free, in which case the superposition angles, $\theta_R$ and $\theta_L$, are set to the same value due to symmetry. The average number of photons $\langle n \rangle$ per party in the TMSV state is fixed at 0.5. In Fig. \ref{Results:Figure:NegVsTheta} we plot the negativity as a function of the angle, $\theta_R=\theta_L$, for a different  number of atomic qubits at each register. We clearly observe a strong dependency on the superposition angle. \\

\begin{figure*}
\centering
\includegraphics[width=2\columnwidth]{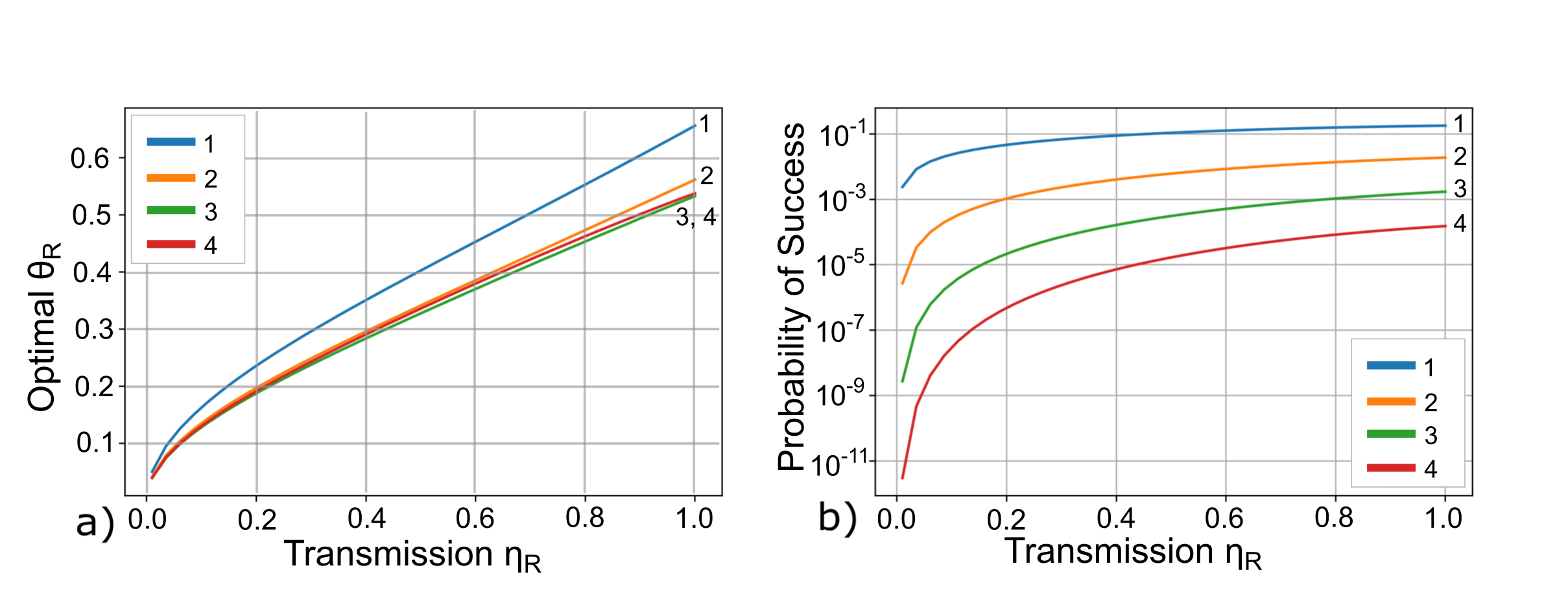}
\caption{\textit{\textbf{a:} The choice of $\theta_R$ (in radians) that maximizes the negativity of the two registers for a given loss in channel ch$_R$. We have fixed $\langle n \rangle$ at 0.5. The legend and plot label indicates the number of qubits in each register.  \textbf{b:} We plot the probability of the measurements at all the photodetectors succeeding, at the optimal value of $\theta_R$. The legend and plot label indicates the number of qubits in each register.}}
\label{Results:Figure:ThetaProbVsEta}
\end{figure*}

\begin{figure}
\centering
\includegraphics[width=1\columnwidth]{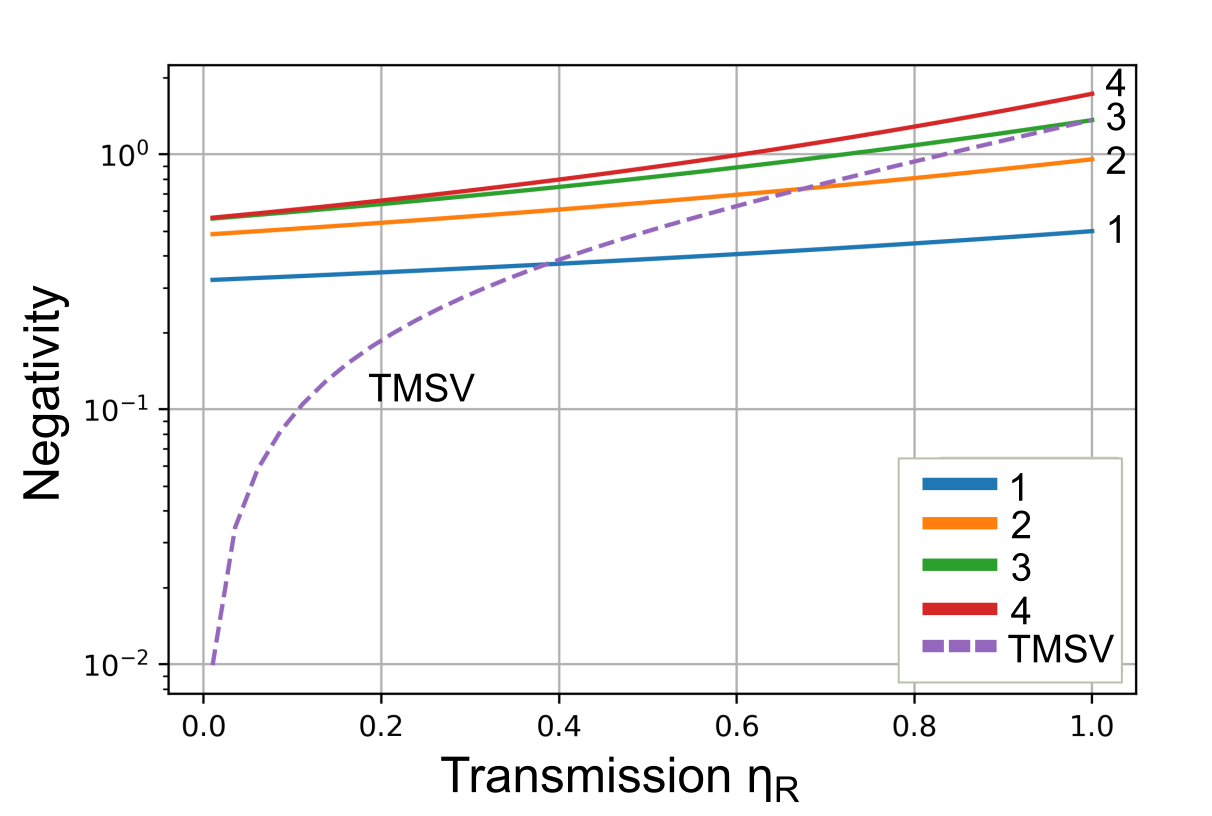}
\caption{\textit{We plot how the negativity depends on the transmission of the channel connecting the registers. The negativity is computed at the optimal value of $\theta_R$ shown in Fig. \ref{Results:Figure:ThetaProbVsEta}a. The average number of photons emitted by the TMSV source $\langle n \rangle$ into each channel is fixed at 0.5. The plot label and legend indicates the number of qubits in each register.}}
\label{Results:Figure:NegVsEta}
\end{figure}

We then fix $\theta_L$ at the value corresponding to the largest negativity, as inferred from Fig. \ref{Results:Figure:NegVsTheta}, and lower the transmission of the right channel (ch$_\texttt{R}$). We allow $\theta_R$ to change and find the angle that maximizes the negativity at different transmissions. The result can be seen in Fig. \ref{Results:Figure:ThetaProbVsEta}a. We find that as the channel transmission, $\eta_R$, is reduced, the angle $\theta_R$ must be changed to maximize the negativity. This can be understood from the fact that a low transmission reduces the probability that photons arrive at the right register. This in turn implies, that a successful measurement outcome at the photodetectors, was entirely due to light emitted from bright state qubits at the register. This lowers the negativity of the two registers since they approach a separable state. However, this effect can be counteracted by lowering the probability that the qubits are in the bright state, which is exactly what is done by lowering $\theta_R$.\\
The probability that the measurement outcomes at the photodetectors correspond to successful entanglement sharing, at the optimal value of $\theta_R$, is shown Fig. \ref{Results:Figure:ThetaProbVsEta}b. We observe that the probability of success decreases exponentially in the number of qubits in the registers, and super-exponentially for decreasing transmission. The super-exponential decrease in the probability of success, is caused by the scheme compensating for a low $\eta_R$ by lowering $\theta_R$. This implies that our entanglement sharing scheme will be practically infeasible at low transmissions and for registers with many qubits.\\ 
In Fig. \ref{Results:Figure:NegVsEta} we show how the negativity of the state shared by the registers depend on the transmission of channel ch$_\texttt{R}$. The negativity is computed at the optimal value of $\theta_R$. For reference we plot the negativity of the TMSV state used to share entanglement between the registers. We observe that for low transmission, the registers have a higher negativity than the TMSV state. This is caused by the noiseless amplification process. Of course, this comes at a cost of probability, with experiments performed at low transmissions ($\eta_R$) having a very low probability of success. On the other hand, when the transmission is high and the registers comprise only one or two qubits, the negativity is in fact decreased by the noiseless amplification process. This is due to the truncation of the Hilbert space, and this effect is mitigated by increasing the number of qubits in the registers.\\
\section{Connecting the Segments via Deterministic Entanglement Swapping}
\label{Analysis:section:entSwap}
\begin{figure}
\centering
\includegraphics[width=1\columnwidth]{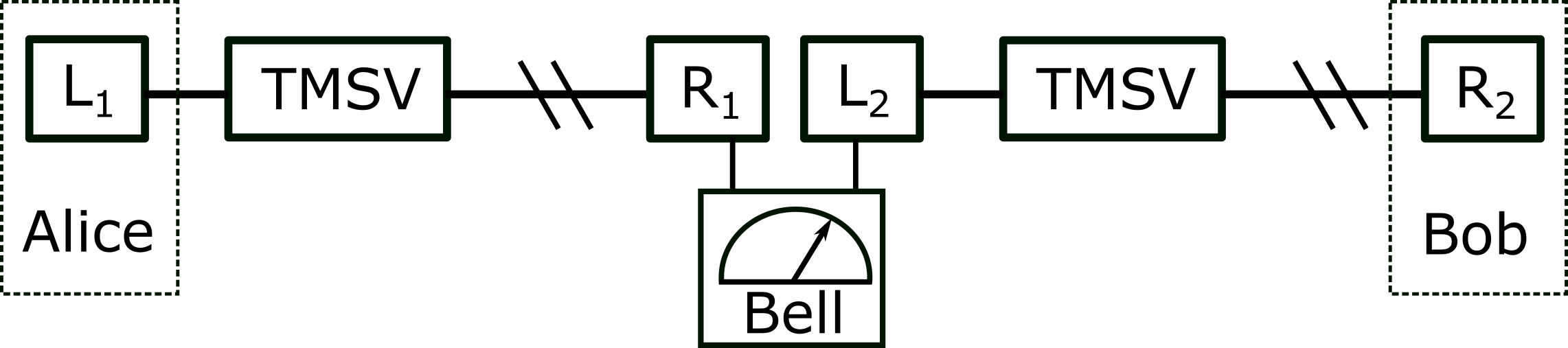}
\caption{\textit{We generate two pair of entangled registers and label them as $L_1$, $R_1$, $L_2$, and $R_2$. We assume we can perform a joint measurement on register $R_1$ and $L_2$.}}
\label{Analysis:Figure:Swap}
\end{figure}

\begin{figure*}
\centering
\includegraphics[width=2\columnwidth]{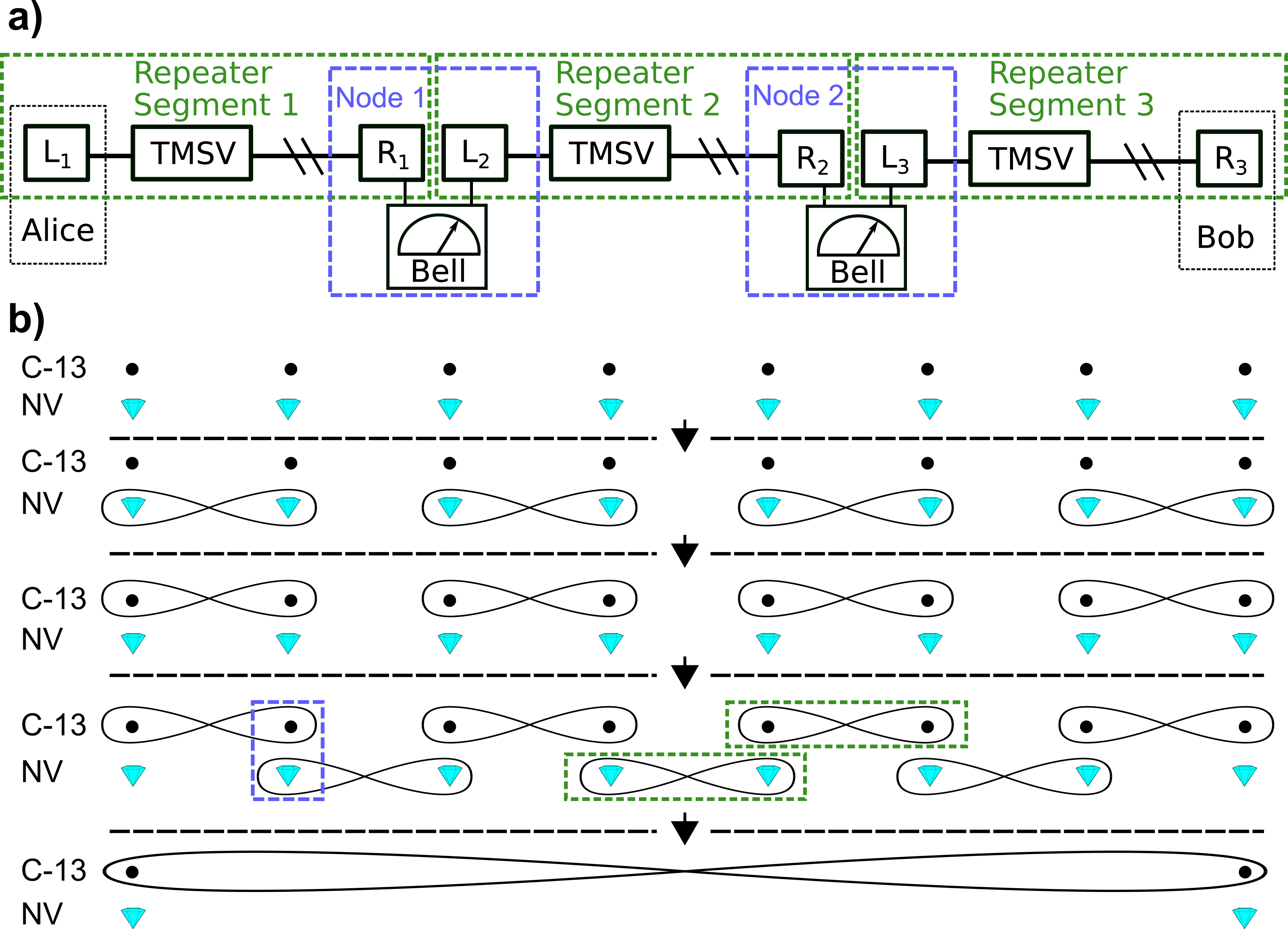}
\caption{\textit{\textbf{a}: We refer to a pair of registers connected by a TMSV state as a repeater segment. Pairs of registers capable of undergoing an entanglement swap forms a repeater node. We may combine repeater segments sequentially to form a repeater, here shown with 3 repeater segments connecting Alice and Bob. \textbf{b}: If we interpret the two qubits in a given node as an NV-center and a nearby coupled $^{13}$C nuclear spin, then the repeater chain must run step-wise as sketched. The involved steps are separated by a dotted line and an arrow, moving from top to bottom. Entanglement between two qubits is indicated by an enveloping loop. We highlight two repeater segments with a green dotted box, and a node in a blue dotted box. In the final step, entanglement swapping is performed at all the nodes, generating an entangled state between Alice and Bob.}}
\label{Analysis:Figure:Swap3}
\end{figure*}

Having established that the superposition angle $\theta$ can be used to increase the negativity between registers, we investigate the possibility of using these registers as a memory unit in a quantum repeater. We will focus our analysis on the case of 1 qubit per register. This case is the most relevant considering current technological limitations.
We now show how entanglement swapping between single-qubit registers is performed. Suppose we have 4 registers, as shown in Fig. \ref{Analysis:Figure:Swap}, pairwise entangled in the state $\rho$ given by Eq. \ref{NoError:Equation:TracedDensityLoss_Analysis} with $N=1$. The total state $\Omega$ is then a product of two of such states $\Omega = \rho \otimes \rho$.
We label the registers as $L_1$, $R_1$, $L_2$, and $R_2$. We assume we can perform a deterministic Bell measurement on the registers $R_1$ and $L_2$.
We may join several pairs of registers in series to form a repeater array as shown in Fig. \ref{Analysis:Figure:Swap3}a.\\
We propose that a repeater node (eg. $R_1$ and $L_2$) could consist of two closely situated (coupled) NV centers, on which we can perform a joint Bell measurement, and we will analyze the repeater based on this assumption. However, we note that the qubits making up a repeater node could also be realized as the electronic spin of a NV center and the spin of a nearby $^{13}$C atom coupled to the NV center. In this case, the repeater protocol would have to be realized step-wise. Eg. referring to Fig. \ref{Analysis:Figure:Swap3}a, repeater segment 1 would establish entanglement between the NV centers at Alice and node 1, and the entangled state would then be transferred to $^{13}$C atoms at Alice and node 1. These nuclear qubits make up $L_1$ and $R_1$. Simultaneously with this, entanglement would be generated between $^{13}$C atoms at node 2 and at Bob using repeater segment 3. These nuclear qubits would in turn make up $L_3$ and $R_3$. Entanglement is then shared between NV centers at node 1 and node 2 using repeater segment 2, these electronic qubits make up $L_2$ and $R_2$. A swap is then performed on the NV center and $^{13}$C nuclear spin at both node 1 and 2, thereby generating an entangled state between Alice and Bob's $^{13}$C nuclear spins. The idea is sketched for a longer repeater in Fig. \ref{Analysis:Figure:Swap3}b.\\
\begin{widetext}
Performing entanglement swapping at all the nodes, we find the normalized density matrix after $s$ swaps:
\begin{align}
    \rho_s = \left[2 + (s+1) (\eta^{-1}-1)  \tan(\theta_R)^2  \right]^{-1} \begin{pmatrix}
    1   & 0 & 0 &  \left(-e^{i \phi}\right)^{s+1} \\
    0 &  (s+1) (\eta^{-1}-1)  \tan(\theta_R)^2 & 0 & 0 \\
    0 & 0 & 0 & 0 \\
    \left(-e^{-i \phi} \right)^{s+1}    & 0 & 0 & 1 \\
    \end{pmatrix}
\end{align}
\end{widetext}
Here we assumed that all $s$ Bell measurements projected onto the state $\ket{\psi} = \frac{1}{\sqrt{2}} \left( \ket{0_{R_n}0_{L_{n+1}}} + \ket{1_{R_n}1_{L_{n+1}}} \right)$. The derivation can be found in appendix A. We note that the matrix element corresponding to loss, $\left( (s+1) (\eta^{-1}-1)  \tan(\theta_R)^2 \right)$, grows linearly in the number of swaps, and will dominate after many swaps. Of course, other Bell measurement outcomes than the one considered here will occur. In our reported results we sample swaps fairly according to the probability at which they occur.

\section{Performance of the Quantum Repeater}
\label{Analysis:section:results}
\begin{figure}
\centering
\includegraphics[width=1\columnwidth]{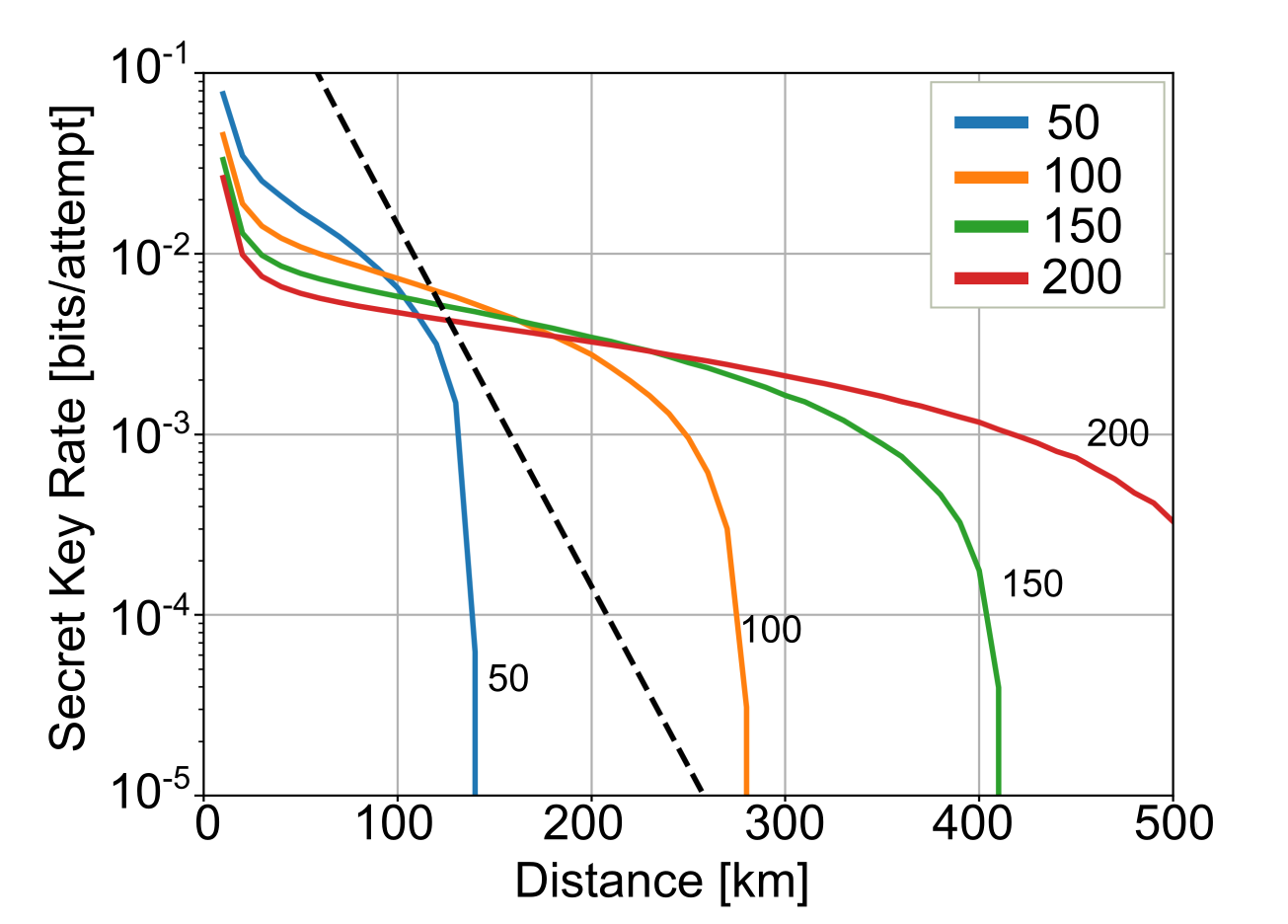}
\caption{\textit{Plot of the secret key rate obtained with 1 qubit per register against the distance between the two participants attempting to share a secret key. Alice is the reconcilliator. The secret key rate is computed for different number of allowed attempts $A$ (legend and plot label). We assume a loss of 0.2 db/km. Each curve is an average over 15 calculations, where each calculation might differ due to the Bell measurement outcomes realized at each swap. The separation of two registers, forming a repeater segment, is 10 km. For reference we plot the PLOB bound \cite{pirandola17} as a dotted black line.}}
\label{Results:Figure:SKsvsDist}
\end{figure}

Having described the construction of the entire quantum repeater scheme, we will now discuss its performance in terms of its ability to generate a secret key between two parties. We will assume that Alice is the reconcilliator. We will also analyze the possibility of violating a Bell inequality, which will enable device independent quantum key distribution (DI-QKD). 
As is derived in Appendix A, when we decrease $\theta_R$ we increase the purity of the state shared by the single qubit registers, however this comes at a cost of a lower probability of success. In a realistic scenario, the experimenter has a finite number of attempts to set up her repeater channel. If the scheme has not succeeded within this number of attempts it might be impractical to use the scheme for sharing secret keys, due to the long waiting time. We take this into account by defining some number of attempts available to the experimenter, $A$. Mathematically we impose the constraint that the average experiment succeeds in $A$ attempts.\\
Let $p$ be the probability that each pair of registers ($L_n,R_n$) successfully generate the shared state given in Eq. \ref{NoError:Equation:TracedDensityLoss_Analysis}. In order for the whole repeater array to succeed in $A$ attempts on average, then $p$ must necessarily be related to $A$. The exact relation is given in Appendix C Eq. \ref{AppC:Equation:fixP}. We determine $p$ numerically from $A$ and insert it into Eq. \ref{Analysis:Equation:thetaR} so that we may determine the optimal values of $\theta_R$, $\theta_L$ and $\langle n \rangle$. We then compute the secret key rate for various values of $A$. An expression for the secret key rate is derived in Appendix B and C, and given by Eq. \ref{Analysis:secretKeyRate}.
The secret key rate can be seen in Fig. \ref{Results:Figure:SKsvsDist} as a function of distance (assuming a fiber loss rate of 0.2dB/km). Our calculations imply that the proposed setup, under the assumed idealizations, might beat the point-to-point capacity bound (also known as the PLOB bound \cite{pirandola17}) at roughly 130 km. However, one should keep in mind that the repeater requires extensive two-way classical communication between segments, and key exchange is expected to be slow. \\

In order to compute the key rates presented in Fig. \ref{Results:Figure:SKsvsDist}, we have numerically optimized a number of repeater parameters, including the length of a single segment, the mean photon number of the TMSV, and the angles, $\theta_L$ and $\theta_R$.
The length of a repeater segment was set to 10 km. This distance was found to be optimal as revealed by the scans shown in Appendix D Fig. \ref{Results:Figure:optSep}. Note that the performance only varies weakly with the segment length. The optimal value of $\langle n \rangle$ as a function of the distance is shown in Appendix D Fig. \ref{Results:Figure:AvgvsDist}. The optimal values of $\theta_L$ and $\theta_R$ are also shown in Appendix D, in Fig. \ref{Results:Figure:thetaLvsDist} and Fig. \ref{Results:Figure:thetaRvsDist} respectively.\\

\subsection{Bell Inequality Violation and Device-Independent QKD}
\begin{figure*}
\centering
\includegraphics[width=2\columnwidth]{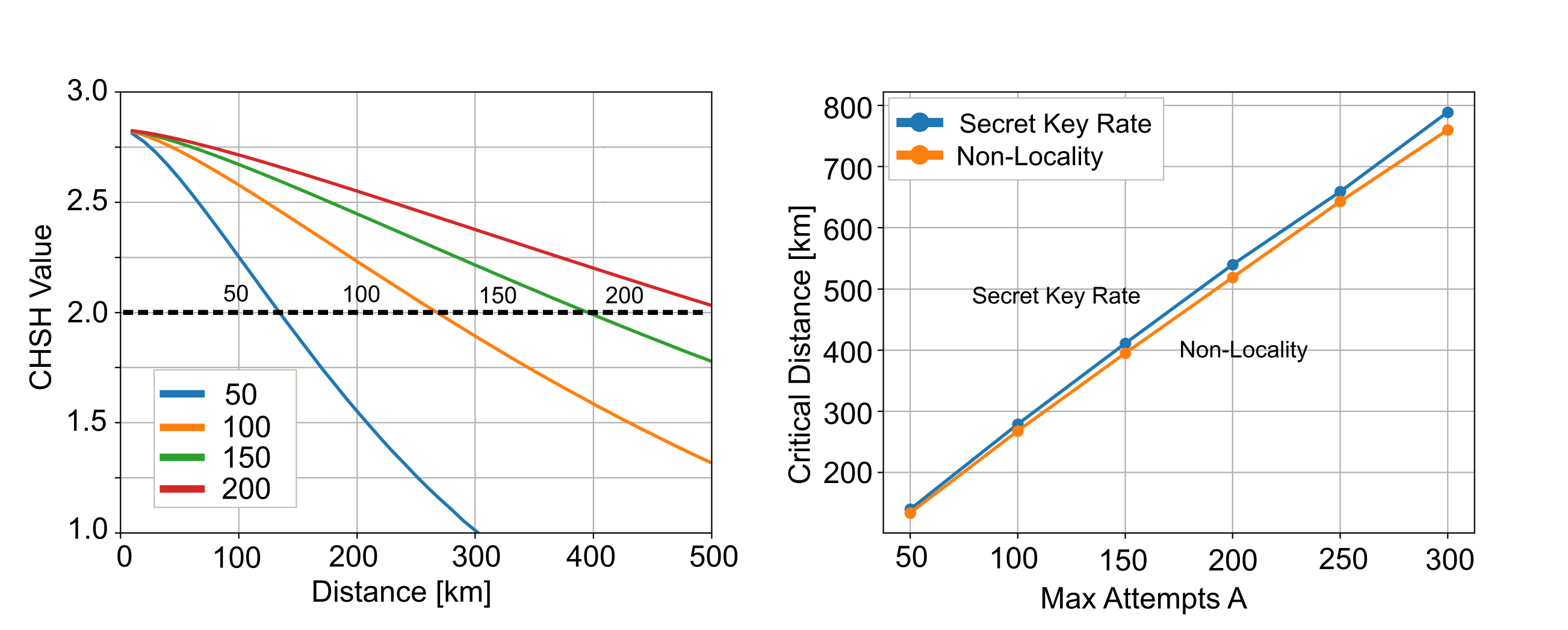}
\caption{\textit{\textbf{Left:} Plot of the CHSH value against the distance between the end points of the repeater. A CHSH value above 2 is inconsistent with a local model. The CHSH value is computed for different number of allowed attempts $A$ (legend and plot label). Each curve is an average over 15 calculations. \textbf{Right:} We plot the critical distance at which the secret key rate vanishes, and the distance at which the CHSH value is equal to 2, against the number of allowed attempts $A$. We observe a nearly linear dependence on $A$ and the curves are nearly identical.}}
\label{Results:Figure:breakDist}
\end{figure*}

\begin{figure*}
\centering
\includegraphics[width=2\columnwidth]{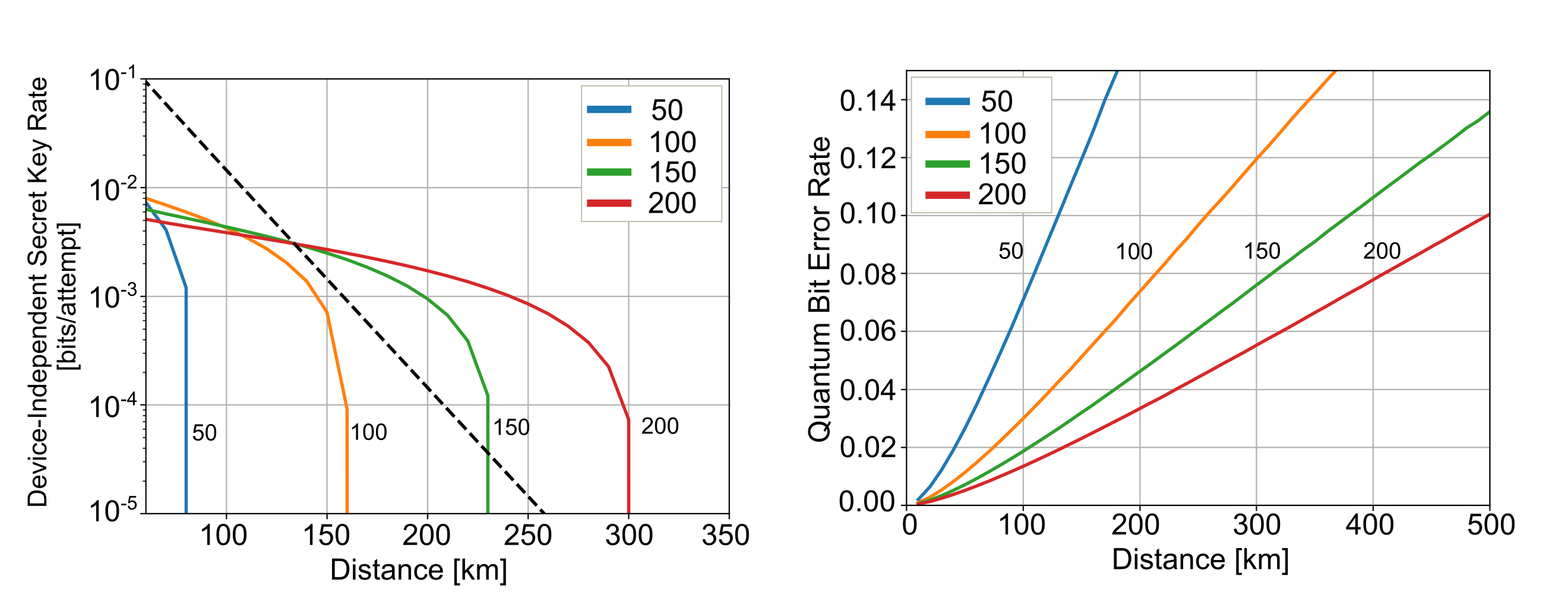}
\caption{\textit{ \textbf{Left:} Device-independent secret key rate against the distance between the end points of the repeater. The rate is computed using Eq. \ref{Results:Equation:DIQKD}. The PLOB bound is drawn as a black dotted line. \textbf{Right:} Quantum bit error rate ($Q$) against the distance. The quantum bit error rate is in our case defined as the probability that Alice and Bob get different outcomes given that they both measure $\sigma_x$, that is  $Q=P(a \neq b|10)$. The legend and plot label indicates the value of $A$.}}
\label{Results:Figure:DIQKD}
\end{figure*}

Device independence represents an ultimate level of security where minimal trust is placed in the implementation of the QKD protocol. A prerequisite for a device independent proof of security is that Alice and Bob (the end points of the repeater) can violate a Bell inequality with their shared 2 qubit state, and that the violation coincide with what they expect based on the quality of the channel in use \cite{vazirani2014,barrett05}. We follow the device independent protocol presented in \cite{pironio09}. We note that Alice is the reconcilliator in our scheme. Alice measures one of the operators
\begin{align}
    M_A^{(1)} &= \sigma_x , \nonumber \\
    M_A^{(2)} &= \sigma_z .
\end{align}
Whereas Bob measures one of the operators
\begin{align}
    M_B^{(0)} &= \sigma_x , \nonumber \\
    M_B^{(1)} &= (\sigma_x + \sigma_z)/\sqrt{2} , \nonumber \\
    M_B^{(2)} &= (\sigma_x - \sigma_z)/\sqrt{2} .
\end{align}
A key can be extracted when Alice happens to measure $M_A^{(1)}$ and Bob happens to measure $M_B^{(0)}$. If Bob measures either $M_B^{(1)}$ or $M_B^{(2)}$, then these measurement outcomes are announced and compared with Alice's measurement outcomes. From this comparison one can compute the value of the CHSH inequality,
\begin{align}
    S = &\langle M_A^{(1)} M_B^{(1)} \rangle + \langle M_A^{(1)} M_B^{(2)} \rangle \nonumber \\ &+ \langle M_A^{(2)} M_B^{(1)} \rangle - \langle M_A^{(2)} M_B^{(2)} \rangle \leq 2.
    \label{Resuts:Equation:CHSH}
\end{align}
A requirement for violating the CHSH inequality is that Alice and Bob keep track of what swaps occurred in the repeater, and perform appropriate corrections to the shared quantum state. In Fig. \ref{Results:Figure:breakDist} to the left we plot the CHSH value $S$ against the distance between Alice and Bob. We note that the critical distance, where $S$ drops below the classical bound of 2, is similar to the distance at which the secret key rate (Fig. \ref{Results:Figure:SKsvsDist}) vanishes for the same value of $A$. To investigate this connection further, we determine the distance at which the secret key rate vanishes for various values of $A$, and compare it with the distance at which the CHSH value drops below 2. The two resulting curves are shown as a function of $A$ in Fig. \ref{Results:Figure:breakDist} to the right. We note that both curves exhibit a nearly linear dependence on $A$, and that the two curves nearly coincide.\\
We compute the device-independent secret key as \cite{pironio09}
\begin{align}
    r \geq 1 - h(Q) - h\left(\frac{1+\sqrt{(S/2)^2-1}}{2}\right),
    \label{Results:Equation:DIQKD}
\end{align}
and normalize by the required number of attempts to set up the repeater. The rate in Eq. \ref{Results:Equation:DIQKD}, while first derived for collective attacks, by entropy accumulation also holds asymptotically for coherent attacks \cite{arnon-friedman18}. The quantum bit error rate $Q$ (QBER), is defined as the probability that Alice and Bob get different measurement outcomes given that they both measure $\sigma_x$, so $Q=P(a \neq b|10)$. We have introduced the binary entropy function $h(x)$. The computed device-independent key rate can be seen in Fig. \ref{Results:Figure:DIQKD} to the left. To the right we show the corresponding values of $Q$. The device-independent key rate appear to be more sensitive to loss than the regular key rate (Fig. \ref{Results:Figure:SKsvsDist}), and as a result, vanishes at shorter distances.

\subsection{Robustness of the Scheme}
\begin{figure*}
\centering
\includegraphics[width=2\columnwidth]{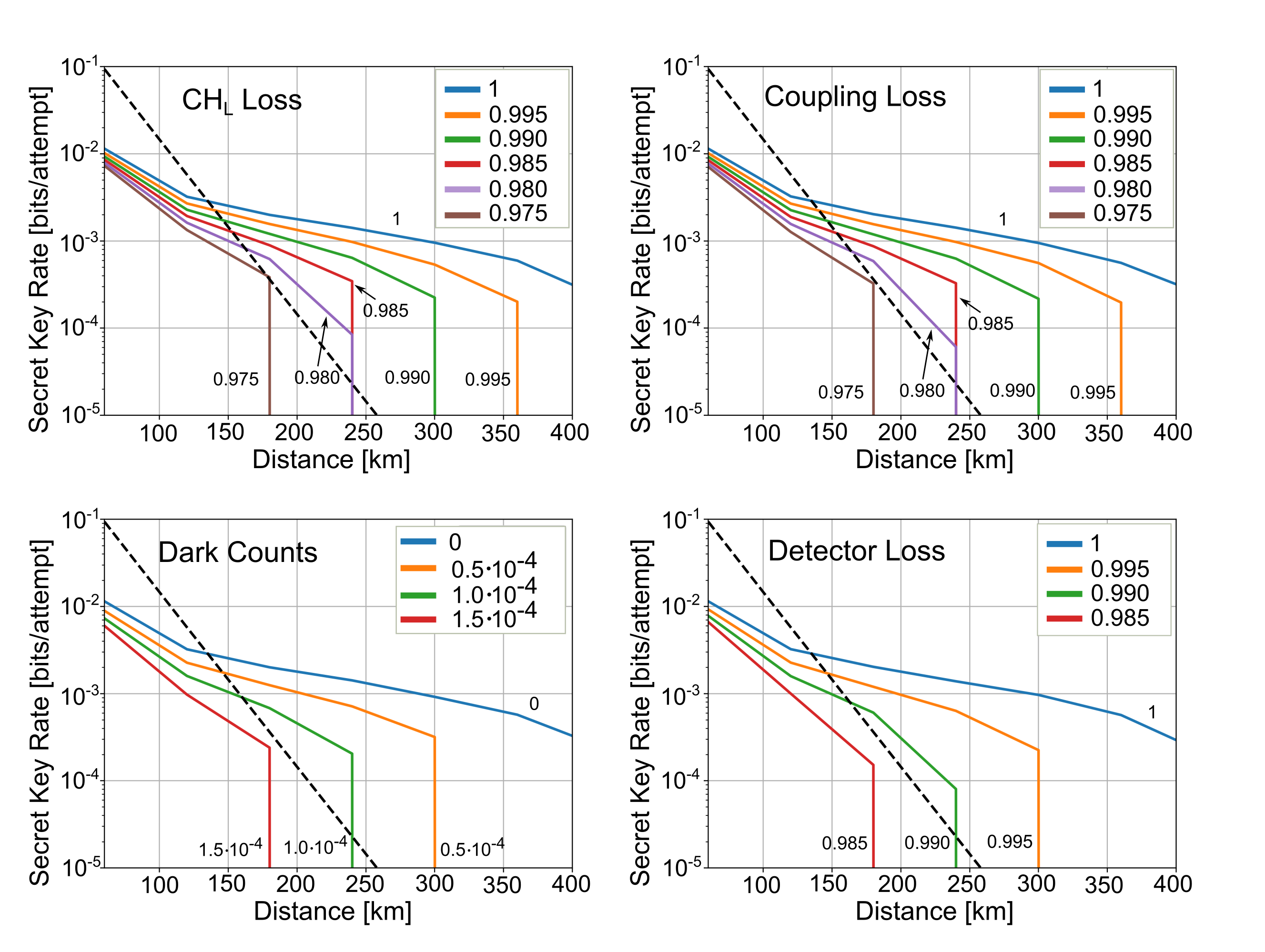}
\caption{\textit{We plot the secret key rate with varying imperfections in the repeater. The distance between repeater nodes is 60 km, and $A$ is close to 500 for all plots. The plots are jagged due to the step size along the x-axis being 60 km.  The PLOB bound is drawn as a black dotted line. \textbf{Top left:} We vary the transmission of the left channel (ch$_L$). The assumed transmission of ch$_L$ is shown in the  legend and plot label. \textbf{Top right:} We vary the transmission when coupling the emission of the NV center to a fiber. The assumed transmission is shown in the legend and plot label.  \textbf{Bottom left:} We vary the probability of a dark count at the detectors $T$ and $B$. The assumed probability is shown in the legend and plot label. \textbf{Bottom right:} We vary the loss of detectors $T$ and $B$. The assumed fraction of light collected by the detectors, is shown in the legend and plot label.}}
\label{Results:Figure:errorModel}
\end{figure*}

We then investigate the robustness of the scheme toward various sources of error. We consider the following 4 errors: Loss in the left channel (ch$_\texttt{L}$), loss when coupling the emission of the NV center to a fiber (channels $f$), dark counts at the detectors $T$ and $B$, and finally, loss in the detectors $T$ and $B$. It is difficult to obtain an analytic expression for the state when including these errors, due to the large number of sums involved. Therefore we turn to a numerical simulation using a custom Python module given on github \cite{github}. We find that when including these errors in our model, it is advantageous to increase the distance of a repeater segment to 60 km, while also increasing $A$ to 500. By increasing the length of a repeater segment, we decrease the required number of segments, and therefore the number of errors. Note that $A$ will vary slightly as we change the error rates, and all key rates are normalized appropriately. The secret key rates obtained from our simulation can be seen in Fig.  \ref{Results:Figure:errorModel}, from which one can gauge the sensitivity of the repeater toward various sources of error. Our calculations indicate, that in order to beat the PLOB bound, loss in channel ch$_L$ should be kept below about one percent. Likewise, coupling loss from the NV center to the fiber should also be kept below about one percent. The loss in the detectors should be less than half a percent, and the probability of a dark count during a measurement should be less than $0.5 \cdot 10^{-4}$ ($0.5 \cdot 10^{-2}$ percent).
We will not in the present work investigate the robustness of the scheme against noise and decoherence in the qubit memories. However, we expect that the probability of a bit flip must be kept below one percent. We base this expectation on the fact that coupling loss and loss in the channel ch$_L$, has an effect on the quantum state which strongly resemble a bit flip error.
\section{Conclusion}
We have analyzed a protocol for generating entanglement between a pair of multi-qubit registers, where entanglement is shared by distributing two-mode squeezed vacuum states followed by noiseless amplification using quantum scissor operations with atomic qubits. Underlying our analysis is the assumption that the qubits can occupy a bright and a dark state. With this in mind, we propose that these registers could be physically realized using NV centers in diamond. We found that entanglement can be increased between the registers by purifying the shared state via tuning of the angles $\theta_L$ and $\theta_R$, with $\sin(\theta_{L})$ ($\sin(\theta_{R})$) being the amplitude corresponding to qubits in the left (right) register being in the bright state. We found that with a single qubit per register it is possible to use the proposed protocol in a repeater, capable of beating the PLOB bound at around 130 km, under ideal conditions. We then gauged the sensitivity of the scheme against various sources of error. We found that in order to beat the PLOB bound, loss of emission from the NV centers should be kept below 1 percent, and loss in the channel ch$_L$ should likewise be below 1 \%. Loss in the detectors should be kept below 0.5 \%, and the probability of a dark count during a measurement should be kept below $0.5 \cdot 10^{-2}$ \%. We computed the value of the CHSH inequality for the analyzed setup, and found that at the distance where the secret key rate vanishes, it is also possible to construct a local hidden variable model for the measurement outcomes. Finally, using the computed CHSH value and the quantum bit error rate, we bounded the device-independent secret key rate of the repeater for a particular protocol.
\section{Acknowledgment}
We acknowledge the support of the Danish National Research Foundation through the Center for Macroscopic Quantum States (bigQ, DNRF0142)." 
\clearpage
\bibliography{bibliography}
\clearpage
\section{Appendix}
\subsection{Generating Entangled Registers}
\begin{figure*}
\centering
\includegraphics[width=2\columnwidth]{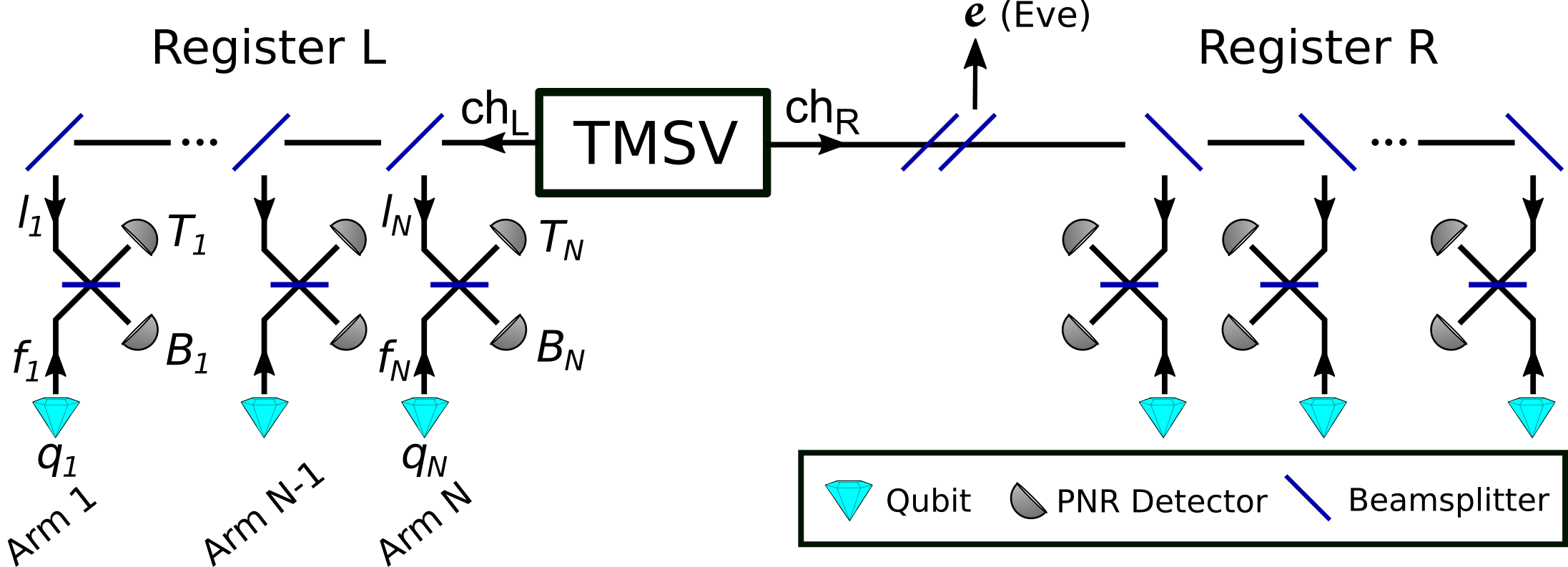}
\caption{\textit{Layout of the entanglement-sharing scheme investigated in this work.}}
\label{Analysis:Figure:LayoutA}
\end{figure*}

We initialize the optical channels (ch$_\texttt{L}$ and ch$_\texttt{R}$) into a TMSV state:
\begin{equation}
	\ket{\psi} = \sum_{n=0}^{\infty} c_n \ket{n}_L\ket{n}_R
\end{equation}
The subscript $L$ and $R$ indicate whether we are referring to the left part of the state or the right part.
The amplitudes are given by:
\begin{equation}
	c_n = (-e^{i \phi})^n  \sqrt{\frac{\langle n \rangle^n}{(1+\langle n \rangle )^{n+1}}}
\end{equation}
Where $\langle n \rangle$ is the average number of photons in each arm of the TMSV state. In this work we consider the case where the phase $\phi$ is set to 0. Focusing on the left side of Figure \ref{Analysis:Figure:LayoutA}. We split up the optical mode on $N$ beamsplitters. We intend to split the optical field evenly among the $N$ modes, to do this we use the transmission and reflection coefficients:
\begin{align}
	t_j = \sqrt{\frac{j-1}{j}}, \hspace{1cm} 	r_j = \sqrt{\frac{1}{j}}
\end{align}
Where $j$ is the number of the arm, starting from 1 at the leftmost arm.
As a result, the amplitude operator for the left part of the TMSV, $a_L$, splits into the $N$ arms as:
\begin{equation}
	a_L \rightarrow \sqrt{\frac{1}{N}} \sum_{k=1}^{N} a_k
\end{equation}
$\ket{\psi}$ then transforms as:
\begin{align}
	\ket{\psi} & = \sum_{n=0}^{\infty} c_n \frac{(a_L^\dagger)^n}{\sqrt{n!}}\ket{0}_L\ket{n}_R\\ &\rightarrow \sum_{n=0}^{\infty} c_n \frac{1}{\sqrt{n!}} \left( \sqrt{\frac{1}{N}} \right)^n (a_1^\dagger + a_2^\dagger + \cdots + a_N^\dagger)^n \ket{0}_l\ket{n}_R \nonumber
\end{align}
Note that we also propagated $\ket{0}_L$ to $\ket{0}_l$ with the latter being the empty $l$ modes at the left register. Using the multinomial theorem, we may rewrite the ladder operator product as:
\small
\begin{align}
	&(a_1^\dagger + a_2^\dagger + \cdots + a_N^\dagger)^n \ket{0}_l \nonumber \\ &= \sum_{j_1 + j_2 + \cdots + j_N = n} \frac{n!}{j_1! j_2! \cdots j_N!} \left(a_1^{\dagger}\right)^{j_1} \left(a_2^{\dagger}\right)^{j_2} \cdots \left(a_N^{\dagger}\right)^{j_N} \ket{0}_l \nonumber \\ &= \sum_{j_1 + j_2 + \cdots + j_N = n} \frac{n!\sqrt{j_1!}\sqrt{j_2!} \cdots \sqrt{j_N!}}{j_1! j_2! \cdots j_N!}  \ket{j_1,j_2,\cdots,j_N}_l \nonumber \\ & = \sum_{\mathbf{j}_n} \Omega_{\mathbf{j}_n} \ket{\mathbf{j}_n}_l
\end{align}
\normalsize
Where the final sum $ \sum_{\mathbf{j}_n}$ runs over unique strings $\mathbf{j}_n = (j_1,j_2,\cdots,j_N)$ such that $j_1+j_2+\cdots+j_N = n$. $\Omega_{\mathbf{j}_n}$ is the prefactor given by:
\begin{align}
    \Omega_{\mathbf{j}_n} = \frac{n!}{\sqrt{j_1!}\sqrt{j_2!} \cdots \sqrt{j_N!}}
\end{align}
Using this notation we then write $\ket{\psi}$ as:
\begin{align}
	\ket{\psi} & =  \sum_{n=0}^{\infty} c_n \frac{1}{\sqrt{n!}} \left( \sqrt{\frac{1}{N}} \right)^n \sum_{\mathbf{j}_n} \Omega_{\mathbf{j}_n} \ket{\mathbf{j}_n}_l \ket{n}_R
	\label{NoErrors:Equation:psi1}
\end{align}
Meanwhile, we initialize the left registry in the state:
\begin{equation}
	\ket{L} = \prod_{k=1}^{N} (\cos(\theta_L)\ket{0}_{q_k} \ket{0}_{f_k} + \sin(\theta_L)\ket{1}_{q_k} \ket{1}_{f_k})
	\label{NoErrors:Equation:L}
\end{equation}
Where the notation $\ket{0}_{q_k} \ket{0}_{f_k}$ indicates that for arm $k$, the qubit ($q_k$) is in the state 0 and the fiber ($f_k$) coupled to the qubit is occupied by 0 photons.\\
We may then write $\ket{L}$ as:
\begin{align}
	\ket{L} &= \sum_{a_1=\{0,1\}} \sum_{a_2=\{0,1\}} \cdots \sum_{a_N=\{0,1\}} \nonumber \\
	& \cos(\theta_L)^{(1-a_1)} \sin(\theta_L)^{a_1} \cdots \cos(\theta_L)^{(1-a_N)} \sin(\theta_L)^{a_N} \nonumber \\ 
	& \ket{a_1}_{q_1}\ket{a_1}_{f_1}\cdots \ket{a_N}_{q_N}\ket{a_N}_{f_N} \nonumber \\
	& = \sum_{\mathbf{a}} \beta(\mathbf{a},\theta_L) \ket{\mathbf{a}}_q \ket{\mathbf{a}}_f
	\label{NoErrors:Equation:Register}
\end{align}
Where we've  introduced a sum over binary lists $\sum_{\mathbf{a}}$ leaving the registry in a superposition of binary states $\ket{\mathbf{a}}_q \ket{\mathbf{a}}_f = \ket{a_1}_{q_1}\ket{a_1}_{f_1}\cdots \ket{a_N}_{q_N}\ket{a_N}_{f_N}$. We've also introduced the parameter $\beta(\mathbf{a},\theta_L)$ to take the angle $\theta_L$ into account. 
We interact $\ket{\psi}$ and $\ket{L}$ using $N$ beamsplitters acting on the $l$ and $f$ modes, we then measure on these modes using the photon resolving detectors $T$ and $B$ for each arm. We assume a particular measurement outcome $(t_k,b_k)_k$, indicating that $t_k$ photons are going to detector $T_k$ and $b_k$ photons are going to detector $B_k$ in arm $k$. Subject to this measurement, the state transforms as:
\begin{align}
	\ket{\psi} \ket{L} &\rightarrow \bra{t_1}_{T_1}\bra{b_1}_{B_1} U_1 \bra{t_2}_{T_2}\bra{b_2}_{B_2} U_2 \nonumber \\ 
	&\cdots \bra{t_N}_{T_N}\bra{b_N}_{B_N} U_N \ket{\psi} \ket{L} 
	\label{NoErrors:Equation:ParticularMeasurement}
\end{align}
Where $U_k$ is the beamsplitter in arm $k$. The bra acting from the left then indicates the projective action of the photon resolving detectors. \\
The right part of the TMSV state is entangled with the left registry qubit $k$ if $t_k + b_k = 1$ since the photon could then have come from either the TMSV state or the bright state of the qubit. Other measurement outcomes tend to yield information about the state of the registry and the TMSV state, and are expected to lower the entanglement, though this assumption could be explored further. In this work we will only accept measurement outcomes where $t_k + b_k = 1$ for all $N$ arms of the registers. \\
To evaluate the above expression, we let the beamsplitters act on the bra. Eg. we examine:
\begin{equation}
	\bra{t_k}_{T_k}\bra{b_k}_{B_k} U_k = (U_k^\dagger \ket{t_k}_{T_k}\ket{b_k}_{B_k})^\dagger
\end{equation}
First let $t_k = 1$ and $b_k = 0$, we assume balanced beamsplitters:
\begin{equation}
	U_k^\dagger \ket{1}_{T_k}\ket{0}_{B_k} = \frac{1}{\sqrt{2}} \left( \ket{0}_{f_k} \ket{1}_{l_k} + \ket{1}_{f_k} \ket{0}_{l_k} \right)
\end{equation}
When we let $t_k = 0$ and $b_k = 1$, we get:
\begin{equation}
	U_k^\dagger \ket{0}_{T_k}\ket{1}_{B_k} = \frac{1}{\sqrt{2}} \left( \ket{0}_{f_k} \ket{1}_{l_k} - \ket{1}_{f_k} \ket{0}_{l_k} \right)
\end{equation}
As required by the reciprocity relations of the beamsplitter. Evidently, measuring  $t_k = 0$ and $b_k = 1$ phase shifts registry qubit $k$ by $\pi$ if it is in the bright state (since the mode $f_k$ is then occupied). We will assume that this phase shift can be corrected experimentally. For simplicity we examine the situation, where for all arms, we obtain the measurements $t_k = 1$ and $b_k = 0$. The measurement described by Eq. \ref{NoErrors:Equation:ParticularMeasurement} then transforms the state as:
\begin{align}
	\ket{\psi} \ket{L} \rightarrow \left(\frac{1}{\sqrt{2}}\right)^N \prod_{k=1}^{N} (\bra{0}_{f_k}\bra{1}_{l_k} + \bra{1}_{f_k}\bra{0}_{l_k})  \ket{\psi} \ket{L}
	\label{Analysis:equation:projection}
\end{align}
Note that we haven't renormalized, and the norm of the state has been reduced.\\
Evidently, if mode $l_k$ contains 1 photon, then mode $f_k$ is vacant, and the qubit state must be $\ket{0}_{q_k}$ (dark). Vice versa if mode $l_k$ contains no photons.
We rewrite the projective measurement as:
\begin{align}
\prod_{k=1}^{N} (\bra{0}_{f_k}\bra{1}_{l_k} + \bra{1}_{f_k}\bra{0}_{l_k}) = \sum_{\mathbf{m}} \bra{\lnot\mathbf{m}}_f \bra{\mathbf{m}}_l
\label{NoErrors:Equation:Measurement}
\end{align}
Where $\mathbf{m}$ is any binary list of length $N$. $\lnot\mathbf{m}$ should be understood as the negation (not) of $\mathbf{m}$. 
Using Eq. \ref{NoErrors:Equation:psi1}, \ref{NoErrors:Equation:Register}, \ref{Analysis:equation:projection}, and \ref{NoErrors:Equation:Measurement}, the result of the measurement is:
\begin{widetext}
\begin{align}
	\ket{\psi} \ket{L} \rightarrow \ket{\alpha} =  \left(\frac{1}{\sqrt{2}}\right)^N \sum_{n=0}^{\infty} \sum_{\mathbf{j}_n}   \sum_{\mathbf{a}} \sum_{\mathbf{m}} c_n \frac{1}{\sqrt{n!}} \left( \frac{1}{\sqrt{N}} \right)^n \beta(\mathbf{a},\theta_L) \Omega_{\mathbf{j}_n} \bra{\lnot\mathbf{m}}_f \ket{\mathbf{a}}_f \ket{\mathbf{a}}_q \bra{\mathbf{m}}_l\ket{\mathbf{j}_n}_l  \ket{n}_R 
\end{align}
\end{widetext}
Since $\mathbf{m}$ is a binary list, then the overlap $\bra{\mathbf{m}}_l \ket{\mathbf{j}_n}_l$ is only non-zero when  the optical input from the TMSV state $\mathbf{j}_n$ is also binary. This implies:
\begin{align}
	\Omega_{\mathbf{j}_n} = n!
\end{align}
Conditioned on the overlap $\bra{\mathbf{m}}_l \ket{\mathbf{j}_n}_l$ being non-zero, we see that $\mathbf{a}$ must be the negation of $\mathbf{j}_n$ so that $\bra{\lnot\mathbf{m}}_f\ket{\mathbf{a}}_f$ is also non-zero. This implies that the state shared between the right TMSV and the left registry after measurement is:
\begin{align}
	\ket{\alpha} = \left(\frac{1}{\sqrt{2}}\right)^N & \sum_{n=0}^{N} \sum_{\mathbf{i}_n} c_n \left( \frac{1}{\sqrt{N}} \right)^n \nonumber \\
	&\beta(n,\theta_L)  \sqrt{n!} \ket{\lnot \mathbf{i}_n}_{L}  \ket{n}_R 
\end{align}
Where $\ket{\lnot \mathbf{i}_n}_{L}$ is the state of the qubits in the left register, and the sum runs over binary lists $\mathbf{i}_n$ of length $N$. $\mathbf{i}_n$ will contain $n$ ones and $N-n$ zeros. $\mathbf{i}_n$ corresponds to the binary distribution of photons in the $l_k$ modes prior to a successful measurement.  The sum over $n$ has been terminated at $N$, since no binary string $\mathbf{i}_n$ exist for $n>N$. We've also used the fact that:
\begin{align}
    \beta(\mathbf{a},\theta_L) = \beta(\lnot \mathbf{i}_n,\theta_L) = \beta(n,\theta_L) = \cos(\theta_L)^{n} \sin(\theta_L)^{N-n}
\end{align}
We may perform the sum over $\mathbf{i}_n$ by introducing the vector:
\begin{align}
    \sum_{\mathbf{i}_n} \ket{\lnot \mathbf{i}_n}_{L} = \binom{N}{n}^{1/2} \ket{\mathbf{I}_{N-n}}_L
\end{align}
Where $\ket{\mathbf{I}_{N-n}}_L$ is normalized and is an even superposition of all binary states containing $N-n$ bright state qubits and $n$ dark state qubits. In terms of these vectors, we find that we may express $\ket{\alpha}$ as:
\begin{align}
	\ket{\alpha} = &\left(\frac{1}{\sqrt{2}}\right)^N \sum_{n=0}^{N}  c_n \left( \frac{1}{\sqrt{N}} \right)^n \beta(n,\theta_L) \nonumber \\
	&\sqrt{n!} \binom{N}{n}^{1/2} \ket{\mathbf{I}_{N-n}}_L  \ket{n}_R 
\end{align}
Comparing with the original TMSV state, we see that the left register transforms a Fock state as:
\begin{align}
    \ket{n} &\rightarrow \left(\frac{1}{\sqrt{2}}\right)^N \left( \frac{1}{\sqrt{N}} \right)^n \beta(n,\theta_L)  \sqrt{n!} \binom{N}{n}^{1/2} \ket{\mathbf{I}_{N-n}} \nonumber \\ &=  \Delta(n,\theta_L)  \ket{\mathbf{I}_{N-n}}
    \label{NoError:Equation:FockTransform}
\end{align}
Where
\begin{align}
	\Delta(n,\theta_L) = \left(\frac{1}{\sqrt{2}}\right)^N \binom{N}{n}^{1/2} \left( \frac{1}{\sqrt{N}} \right)^n    \sqrt{n!} \beta(n,\theta_L)
\end{align}
Using this transform, and assuming no loss in the fiber, we may infer that storing the right part of the TMSV in the right register (see Fig. \ref{Analysis:Figure:LayoutA}), results in the state:
\begin{align}
	\ket{\alpha} = \sum_{n=0}^{N} c_n  \Delta(n,\theta_L) \Delta(n,\theta_R)  \ket{\mathbf{I}_{N-n}}_L \ket{\mathbf{I}_{N-n}}_R
	\label{Appendix:Equation:Final1}
\end{align}
We may then compute the norm of the state of the two registers:
\begin{align}
    \braket{\alpha|\alpha} &= \sum_{m=0}^{N} |c_m|^2 |\Delta(m,\theta_L)|^2 |\Delta(m,\theta_R)|^2 
\end{align}
The probability of successful entanglement sharing between the 2 registries is then this norm multiplied by the number of measurements that would yield an equivalent state. For each arm we have 2 detector outcomes, (0,1) or (1,0), that would yield a state equivalent to the one described above. The probability of successful entanglement sharing is then:
\begin{align}
    P_s = 2^N 2^N \braket{\alpha|\alpha} = 4^N \braket{\alpha|\alpha}
\end{align}

\subsubsection{Performance with Loss}
We initialize the optical channel in a TMSV state:
\begin{equation}
	\ket{\psi} = \sum_{n=0}^{\infty} c_n \ket{n}_L\ket{n}_R
\end{equation}
We send the right part of the TMSV through a lossy channel modelled by a beamsplitter with transmission amplitude $\sqrt{\eta_R}$. The transmission of the fiber is then given by $\eta_R$. Likewise we send the left part of the TMSV state through a lossy channel of transmission $\eta_L$  The channels transforms the state as:
\begin{align}
	\ket{\psi} \rightarrow 
	&\sum_{n=0}^{\infty} \sum_{l_R,l_L=0}^{n} \epsilon_R(n,l_R) \epsilon_L(n,l_L)  c_n \nonumber \\
	&\ket{n-l_L}_L \ket{n-l_R}_R \ket{l_L}_{e_L} \ket{l_R}_{e_R} 
\end{align}
Where $e_R$ is a loss channel. The loss amplitude is given as:
\begin{equation}
	\epsilon_{R/L}(n,l_{R/L}) = \sqrt{\binom{n}{n-l_{R/L}}} \eta_{R/L}^{(n-l_{R/L})/2} (1-\eta_{R/L})^{l_{R/L}/2}
\end{equation}
Inserting the Fock state transform Eq. \ref{NoError:Equation:FockTransform}, we arrive at the register state:
\begin{align}
	\ket{\alpha} = &\sum_{n=0}^{\infty} \sum_{l_L=0}^{n} \sum_{l_R=0}^{n} c_n \epsilon_L(n,l_L) \epsilon_R(n,l_R) \Delta(n-l_L,\theta_L) \nonumber \\ 
	&\Delta(n-l_R,\theta_R)  \Theta(N+l_L-n)\Theta(N+l_R-n) \nonumber \\ 
	&\ket{\mathbf{I}_{N-n+l_L}}_L   \ket{\mathbf{I}_{N-n+l_R}}_R \ket{l_L}_{e_L}  \ket{l_R}_{e_R} 
	\label{NoError:Equation:Lossy}
\end{align}
Where we've introduced the step function:
\begin{align}
    \Theta(x) =     \begin{cases}
        1 & \text{if } x \geq 0\\
        0 & \text{if } x < 0
    \end{cases}
\end{align}
The step function takes into account that when more than $N$ photons reach either registers, then the projective measurement fails.
\begin{widetext}
The corresponding density matrix is:
\begin{align}
	\sigma &= \sum_{n=0}^{\infty} \sum_{m=0}^{\infty} \sum_{l_L=0}^{n} \sum_{k_L=0}^{m} \sum_{l_R=0}^{n} \sum_{k_R=0}^{m} c_n c_m^* \epsilon_R(n,l_R) \epsilon_L(n,l_L) \epsilon_R(m,k_R)^* \epsilon_L(m,k_L)^*  \nonumber \\& \Delta(n-l_L,\theta_L) \Delta(n-l_R,\theta_R) \Delta(m-k_L,\theta_L)^* \Delta(m-k_R,\theta_R)^*
	\nonumber \\& \Theta(N+l_L-n)\Theta(N+l_R-n)\Theta(N+k_L-m)\Theta(N+k_R-m)
	\nonumber \\&  \ket{\mathbf{I}_{N-n+l_L}}_L \bra{\mathbf{I}_{N-m+k_L}}_L   \ket{\mathbf{I}_{N-n+l_R}}_R \bra{\mathbf{I}_{N-m+k_R}}_R \ket{l_L}_{e_L} \bra{k_L}_{e_L} \ket{l_R}_{e_R} \bra{k_R}_{e_R}
\end{align}
We then trace out the loss channels, giving the state:
\begin{align}
	\rho &= \text{Tr}_e \sigma = \sum_{n=0}^{\infty} \sum_{m=0}^{\infty} 
	\sum_{l=0}^{\texttt{min}(n,m)} 
	\sum_{r=0}^{\texttt{min}(n,m)} 
	c_n c_m^* \epsilon_R(n,r) \epsilon_L(n,l) \epsilon_R(m,r)^* \epsilon_L(m,l)^*  \nonumber \\& \Delta(n-l,\theta_L) \Delta(n-r,\theta_R) \Delta(m-l,\theta_L)^* \Delta(m-r,\theta_R)^*
	\nonumber \\& \Theta(N+l-n)\Theta(N+r-n)\Theta(N+l-m)\Theta(N+r-m)
	\nonumber \\&  \ket{\mathbf{I}_{N-n+l}}_L \bra{\mathbf{I}_{N-m+l}}_L   \ket{\mathbf{I}_{N-n+r}}_R \bra{\mathbf{I}_{N-m+r}}_R 
\end{align}

We define the matrix elements:
\begin{align}
    \Lambda(n,m,l,r) &= c_n c_m^* \epsilon_R(n,r) \epsilon_L(n,l) \epsilon_R(m,r)^* \epsilon_L(m,l)^*  \nonumber \\& \Delta(n-l,\theta_L) \Delta(n-r,\theta_R) \Delta(m-l,\theta_L)^* \Delta(m-r,\theta_R)^*
    \nonumber \\& \Theta(N+l-n)\Theta(N+r-n)\Theta(N+l-m)\Theta(N+r-m)
	\label{Analysis:Registers:Equation:Lambda}
\end{align}
Such that we may write the state as:
\begin{align}
    \rho = \sum_{n=0}^{\infty} \sum_{m=0}^{\infty} \sum_{l=0}^{\texttt{min}(n,m)} \sum_{r=0}^{\texttt{min}(n,m)} \Lambda(n,m,l,r) \ket{\mathbf{I}_{N-n+l}}_L \bra{\mathbf{I}_{N-m+l}}_L   \ket{\mathbf{I}_{N-n+r}}_R \bra{\mathbf{I}_{N-m+r}}_R
    \label{NoError:Equation:TracedDensityLoss}
\end{align}
\subsubsection{One Qubit Per Register}
\label{Analysis:SingleQubitCase}
In the following we will analyse the case of one sided loss. Loss is assumed to only occur between the EPR source and the right register. If $N=1$ then we have the density matrix:
\begin{align}
    &\rho = \nonumber \\&\frac{1}{4} \begin{pmatrix}
    |c_1|^2 \eta \cos(\theta_L)^{2}  \cos(\theta_R)^{2} & 0 & 0 &  c_1 c_0^* \sqrt{\eta} \sin(\theta_R) \sin(\theta_L) \cos(\theta_R) \cos(\theta_L) \\
    0 & |c_1|^2 (1-\eta) \cos(\theta_L)^2 \sin(\theta_R)^2 & 0 & 0 \\
    0 & 0 & 0 & 0 \\
    c_1^* c_0 \sqrt{\eta} \sin(\theta_R) \sin(\theta_L) \cos(\theta_R) \cos(\theta_L) & 0 & 0 & |c_0|^2 \sin(\theta_L)^2 \sin(\theta_R)^2 \\
    \end{pmatrix}
    \label{Appendix:Equation:1QubitLossy}
\end{align}

If we use the above state for entanglement swapping via Bell measurements, there is an advantage in keeping the matrix elements $\rho_{11}$ and $\rho_{44}$ identical. If this is not the case, series of swaps will tend to make the state more separable and thereby diminish entanglement.
This constraint implies:
\begin{align}
        |c_1|^2 \eta \cos(\theta_L)^{2}  \cos(\theta_R)^{2} = |c_0|^2 \sin(\theta_L)^2 \sin(\theta_R)^2
\end{align}
This implies a bond between $\theta_R$ and $\theta_L$.
\begin{align}
    \tan(\theta_L)^2 = \tan(\theta_R)^{-2} \eta \frac{|c_1|^2}{|c_0|^2}
    \label{Analysis:Equation:Bond}
\end{align}
Utilizing the bond in Eq. \ref{Analysis:Equation:Bond} we may rewrite the density matrix as:
\begin{align}
    \rho = \frac{1}{4} \frac{\sin(\theta_R)^2 |c_1|^2 |c_0|^2 \eta}{ \tan(\theta_R)^2 |c_0|^2+ \eta |c_1|^2} \begin{pmatrix}
    1   & 0 & 0 &  -e^{i \phi} \\
    0 &  (\eta^{-1}-1)  \tan(\theta_R)^2 & 0 & 0 \\
    0 & 0 & 0 & 0 \\
    -e^{-i \phi}    & 0 & 0 & 1 \\
    \end{pmatrix}
    \label{Analysis:Equation:1QubitLossy-2}
\end{align}
\end{widetext}
We now want to choose the superposition angle $\theta_R$ such that the state is as entangled as possible. 
Evidently we have that if we suppress the loss term, corresponding to matrix element $\rho_{22}$, then the state of the qubits is a maximally entangled state. This suggests that we should make $\tan (\theta_R)^2$ as small as possible.
However, there is one more condition to consider. The probability of the measurements at the photodetectors succeeding, which goes to zero in this limit where $\theta_R \rightarrow 0$.
The probability of success is given by the trace of $\rho$:
\begin{align}
    P(\theta_R,\langle n \rangle;\eta) = \sin(\theta_R)^2 |c_1|^2 |c_0|^2 \frac{ 2 \eta + (1-\eta)  \tan(\theta_R)^2}{ \tan(\theta_R)^2 |c_0|^2+ \eta |c_1|^2}
    \label{Analysis:Equation:ProbabilityThetaR}
\end{align}
Where we take into account that the measurements at the photodetectors can succeed in 4 ways. The optimal choice of $\theta_R$ is then the choice that minimizes $\tan (\theta_R)^2$ subject to the condition that:
\begin{align}
    P(\theta_R,\langle n \rangle;\eta) \geq p
    \label{Analysis:Equation:Epsilon}
\end{align}
Where $p$ is  the minimum probability of success tolerated by the experimental setup. 
We may maximize $P(\theta_R,\langle n \rangle;\eta)$ in $\langle n \rangle$ for a given $\theta_R$ and $\eta$ by choosing the squeezing of the TMSV source such that we have the equality:
\begin{align}
    \langle n \rangle = \sqrt{1-\frac{\eta}{\eta + \tan\left(\theta_R \right)^2}}
\end{align}
Inserting the optimal choice of $\langle n \rangle$ into $P(\theta_R,\langle n \rangle;\eta)$ we obtain the probability  $P(\theta_R;\eta)$. From numerical investigation we find that this probability depends on the angle $\theta_R$ in a complicated manner. However, for values of $|\theta_R|$ below $\sim 0.66$ we have that $P(\theta_R;\eta)$ decreases monotonically in $|\theta_R|$ for any value of $\eta$. In order to minimize $\tan (\theta_R)^2$ we require low values of $\theta_R$, and so we expect to be below this angle. With this consideration, we may infer that the optimal choice of $\theta_R$ is obtained when:
\begin{align}
    P(\theta_R;\eta) = p
    \label{Analysis:Equation:thetaR}
\end{align}
Which can be rewritten as a quartic polynomial equation in $z=\sqrt{1-\frac{\eta}{\eta+\tan(\theta_R)^2}}$:
\begin{align}
    p + 2 p z + & \eta \left[ p -2 \right] z^2 + p \left[ 2 \eta - 2 \right] z^3 \nonumber \\ &+ \left[ -p + (1+p) \eta + \eta^2 \right] z^4 = 0
\end{align}
Which can be solved efficiently numerically and from which we can find $\theta_R$.\\
\begin{widetext}
Given that we've made the optimal choice of $\theta_R$ we can investigate how the state evolves under a sequence of swaps. Suppose we have 4 registers, as shown in Fig. \ref{Analysis:Figure:Swap}, pairwise entangled in the state $\rho$ given by Eq. \ref{Analysis:Equation:1QubitLossy-2}. The total state $\Omega$ is then a product of two of such states $\Omega = \rho \otimes \rho$. Obtaining the particular Bell measurement outcome corresponding to the ket $\ket{\psi} = \frac{1}{\sqrt{2}} \left( \ket{0_{R_1}0_{L_2}} + \ket{1_{R_1}1_{L_2}} \right)$ on registers $R_1$ and $L_2$, we obtain the state:
\begin{align}
    \rho_1 = \frac{1}{2} (\bra{0_{R_1}0_{L_2}} + \bra{1_{R_1}1_{L_2}}) \Omega (\ket{0_{R_1}0_{L_2}} + \ket{1_{R_1}1_{L_2}})
\end{align}
We may evaluate $\rho_1$ by inserting $\rho$ from Eq. \ref{Analysis:Equation:1QubitLossy-2}. We obtain the unormalized state:

\begin{align}
    \rho_1 = \frac{1}{2} \left( \frac{1}{4} \frac{\sin(\theta_R)^2 |c_1|^2 |c_0|^2 \eta}{ \tan(\theta_R)^2 |c_0|^2+ \eta |c_1|^2} \right)^2 \begin{pmatrix}
    1   & 0 & 0 &  \left(-e^{i \phi}\right)^2 \\
    0 &  2 (\eta^{-1}-1)  \tan(\theta_R)^2 & 0 & 0 \\
    0 & 0 & 0 & 0 \\
    \left(-e^{-i \phi} \right)^2    & 0 & 0 & 1 \\
    \end{pmatrix}
\end{align}
Clearly the state is similar in structure to the original state, with the loss term having doubled in size relative to the other matrix elements. Continuing this, then after $s$ swaps we obtain the unnormalized density matrix: 
\begin{align}
    \rho_s = \left(\frac{1}{2} \right)^s \left( \frac{1}{4} \frac{\sin(\theta_R)^2 |c_1|^2 |c_0|^2 \eta}{ \tan(\theta_R)^2 |c_0|^2+ \eta |c_1|^2} \right)^{s+1} \begin{pmatrix}
    1   & 0 & 0 &  \left(-e^{i \phi}\right)^{s+1} \\
    0 &  (s+1) (\eta^{-1}-1)  \tan(\theta_R)^2 & 0 & 0 \\
    0 & 0 & 0 & 0 \\
    \left(-e^{-i \phi} \right)^{s+1}    & 0 & 0 & 1 \\
    \end{pmatrix}
\end{align}
Of course, other Bell measurement outcomes than the one considered here will occur. We will numerically simulate swaps by drawing fairly from the four Bell measurement outcomes.
\end{widetext}

\subsection{Secret Key Rate}
\label{Section:SecretKey}
Having established entanglement between registers we're interested in computing how large a shared secret key might be extractable from the density matrix. We will assume that Eve can perform a collective attack \cite{weedbrook12}. Given that Alice and Bob each measure their registry qubit, the secret information is then simply the mutual information between the observed outcomes ($a$ and $b$) minus the information an eavesdropper might have of the outcome obtained by the reconcilliator $x$ (Alice or Bob) \cite{garcia7}. The secret information is given by the Devetak-Winter formula \cite{devetak05}:
\begin{align}
    K =  \beta I(a:b) - S(x:E) 
    \label{Analysis:secretKeyEq}
\end{align}
Where $\beta$ is the reconcilliation efficiency, that is, how large a part of the mutual information that can be distilled into a shared key. The mutual information between the measurement outcomes obtained by Alice and Bob is simply \cite{barnett9}:
\begin{align}
    I(a:b) = \sum_{a,b} P(a,b) \log_2 \left( \frac{P(a,b)}{P(a) P(b)}  \right)
\end{align}
Where $P(a,b)$ is the probability of obtaining outcomes $a$ and $b$, whereas $P(a)$ and $P(b)$ are the marginal probabilities of obtaining outcomes $a$ and $b$ respectively.
$S(x:E)$ is the Holevo information which upper bounds the information Eve can obtain about the variable $x$ given her measurement $e$ \cite{nielsen10}:
\begin{align}
    I(x:e) \leq S(x:E)
\end{align}
With the capital $E$ indicating the state on which Eve has not yet measured. We will assume that Alice is the reconcilliator $x=a$. The Holevo value is given by
\begin{align}
    S(a:E) = S(\rho_{E}) - \sum_a P(a) S(\rho_E^a) .
\end{align}
Where $\rho_E^a$ is the state held by Eve subject to the condition that Alice measures $a$. Since Bob purifies the state $\rho_E^a$, we have that
\begin{align}
    S(\rho_E^a) &= S(\rho_B^a) .
\end{align}
Furthermore, since Alice and Bob purifies the state held by Eve
\begin{align}
    S(\rho_E) = S(\rho_{AB}) .
\end{align}
So we may compute the Holevo value simply by knowing the state shared by Alice and Bob
\begin{align}
    S(a:E) = S(\rho_{AB}) - \sum_a P(a) S(\rho_B^a) .
\end{align}
To obtain a secret key \textit{rate}, we normalize $K$ by the number of channel uses necessary to generate that secret key. The number of channel uses neccessary will be established in Appendix C. We will be assuming a reconcilliation efficiency $\beta$ of 1, and that Alice and Bob either measure $\sigma_x$ or $\sigma_z$ (e.g. they could use the BB84 protocol). They can then extract a secret key from measurement rounds where their choice of basis coincide. 

\subsection{Trials Needed Before N Repeaters Succeed}
In the previous sections we analyzed a single pair of registers, $L_1$ and $R_1$, and found the probability $p$ with which we successfully generate the state $\rho$ given in Eq. \ref{NoError:Equation:TracedDensityLoss}, with $p$ given by $4^N\text{Tr}\rho$. Given $p$ we will assume that the probability of successfully generating $\rho$ after \textit{exactly} $n$ attempts, follows a geometric distribution, with probability mass function:
\begin{align}
    \text{PMF}(n) = p(1-p)^{n-1}
\end{align}
The corresponding cumulative distribution function, which should be interpreted as the probability that $\rho$ has been established in less than or exactly $n$ attempts, is given by:
\begin{align}
    \text{CDF}(n) = 1-(1-p)^n
\end{align}
Now we will assume that we have a collection of $M$ repeater segments, each repeater segment being a pair of registers, as shown in Fig. \ref{Analysis:Figure:Swap3} a. We now want to compute how many attempts are necessary before all $M$ repeater segments successfully generate the state $\rho$.  The cumulative distribution for $M$ repeaters attempting in parallel is simply the product:
\begin{align}
    \text{CDF}_M(n) = [1-(1-p)^n]^M
\end{align}
The probability of all repeaters having succeeded after exactly $n$ attempts is then:
\begin{align}
    \text{PDF}_M(n) &= \text{CDF}_M(n) - \text{CDF}_M(n-1) \nonumber \\ &=[1-(1-p)^n]^M - [1-(1-p)^{n-1}]^M \nonumber \\
    & = \sum_{s=0}^M \binom{M}{s} (1-p)^{n\cdot s} \left[ 1 - (1-p)^{-s} \right]
    \label{Analysis:Equation:ProbSuccessRep}
\end{align}
A similar formula may be found in \cite{bernardes11}. \\
We then fix $p$ by demanding that the following relation is satisfied:
\begin{align}
    A = \sum_{n=1}^\infty n \cdot \text{PDF}_M(n)
    \label{AppC:Equation:fixP}
\end{align}
Implying that the average experiment succeeds in $A$ attempts. \\
Assuming that a state $\rho_M$ is generated from $M-1$ deterministic entanglement swaps using $M$ repeater segments, and that a secret key $K_M$ can be extracted from this state, we normalize this key by the number of attempts necessary to generate $\rho_M$. This gives us the secret key rate $\mathcal{K}_M$:
\begin{align}
    \mathcal{K}_M = \sum^\infty_{n=1}\frac{K_M}{n} \cdot \text{PDF}_M(n) =  K_M \sum^\infty_{n=1}\frac{\text{PDF}_M(n)}{n}
    \label{Analysis:secretKeyRate}
\end{align}
Where $K_M$ is computed from Eq. \ref{Analysis:secretKeyEq}. 


\subsection{Optimal Parameters - 1 Qubit per Register}
\begin{figure*}
\centering
\includegraphics[width=2\columnwidth]{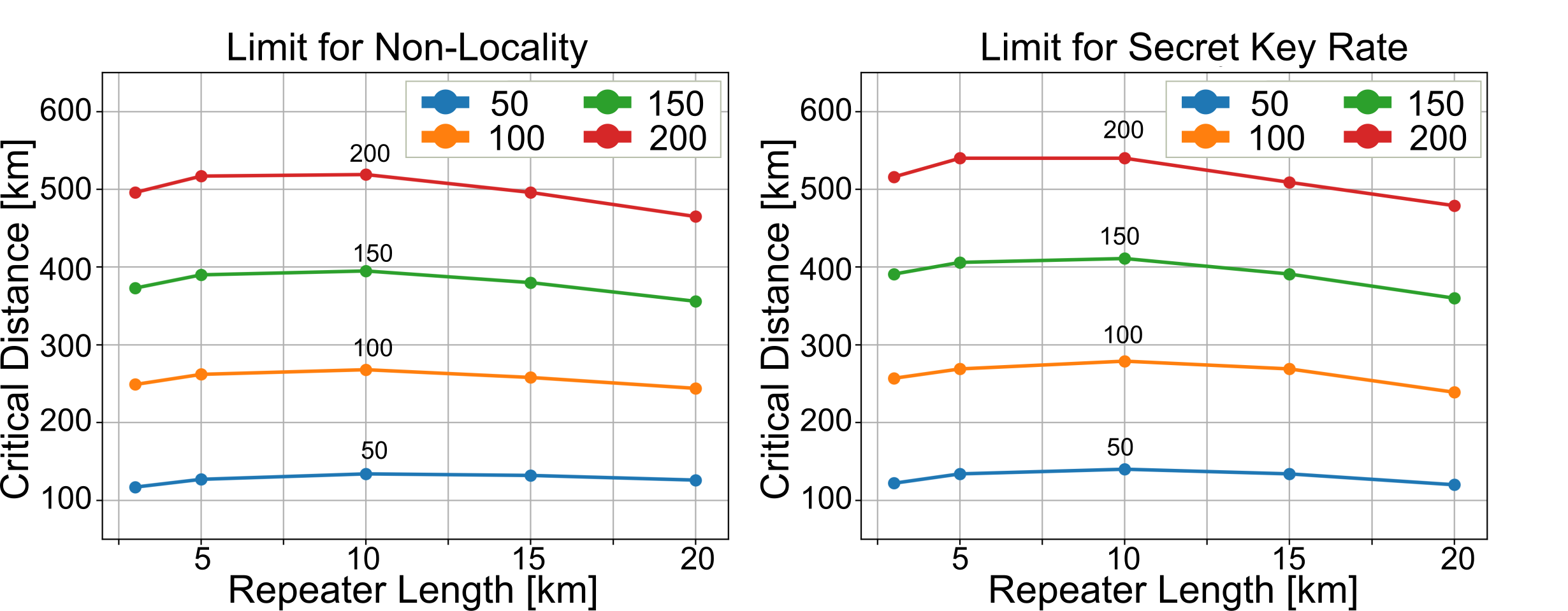}
\caption{\textit{\textbf{Left}: Plot of the distance where no Bell inequality can be violated for various register separations with 1 qubit per register. \textbf{Right}: Plot of the distance where the secret key rate vanishes for various register separations with 1 qubit per register. The legend and plot label indicates the expected number of attempts $A$. }}
\label{Results:Figure:optSep}
\end{figure*}

\begin{figure}
\centering
\includegraphics[width=1\columnwidth]{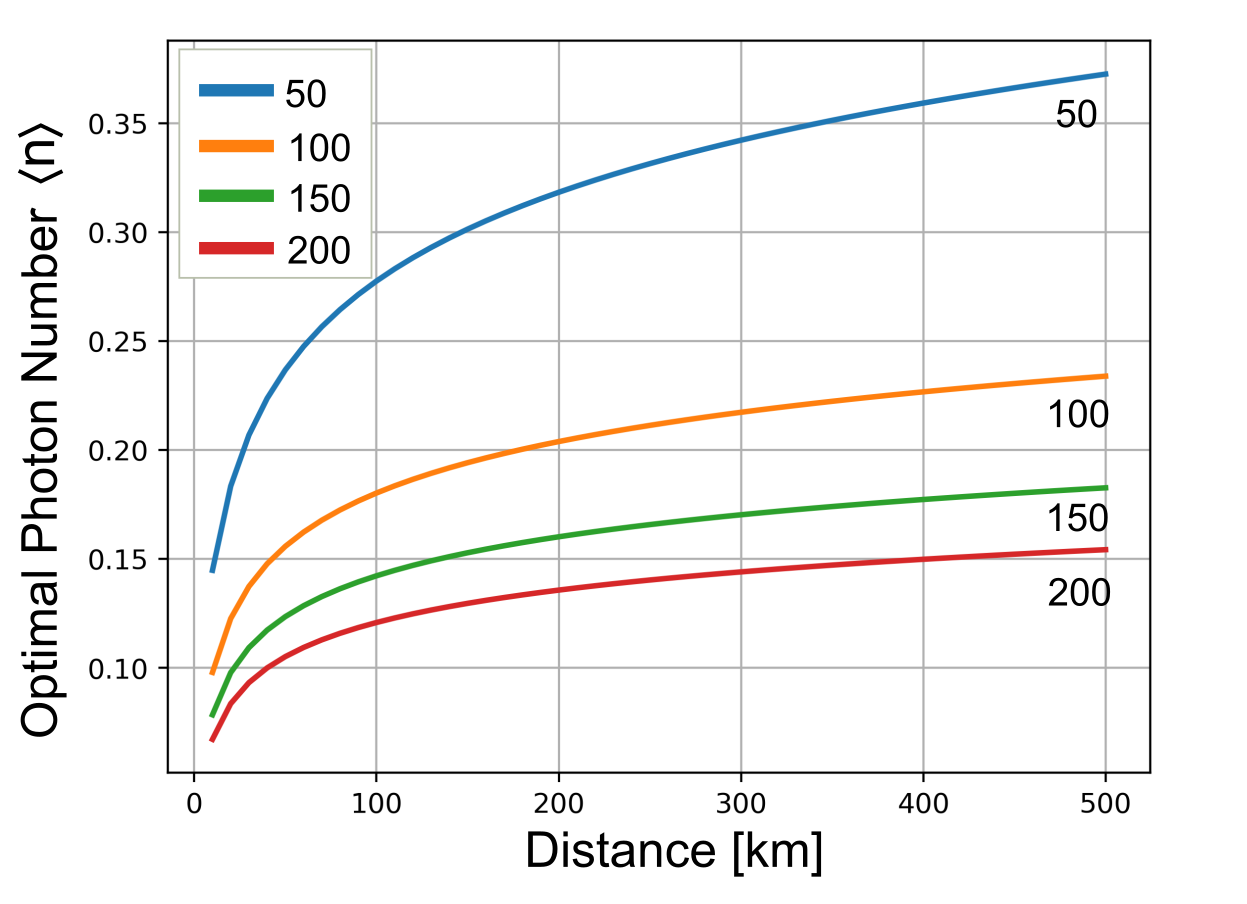}
\caption{\textit{The optimal value of $\langle n \rangle$ for different number of allowed attempts $A$ (legend and plot label) against the distance between the end points of the repeater.}}
\label{Results:Figure:AvgvsDist}
\end{figure}

\begin{figure}
\centering
\includegraphics[width=1\columnwidth]{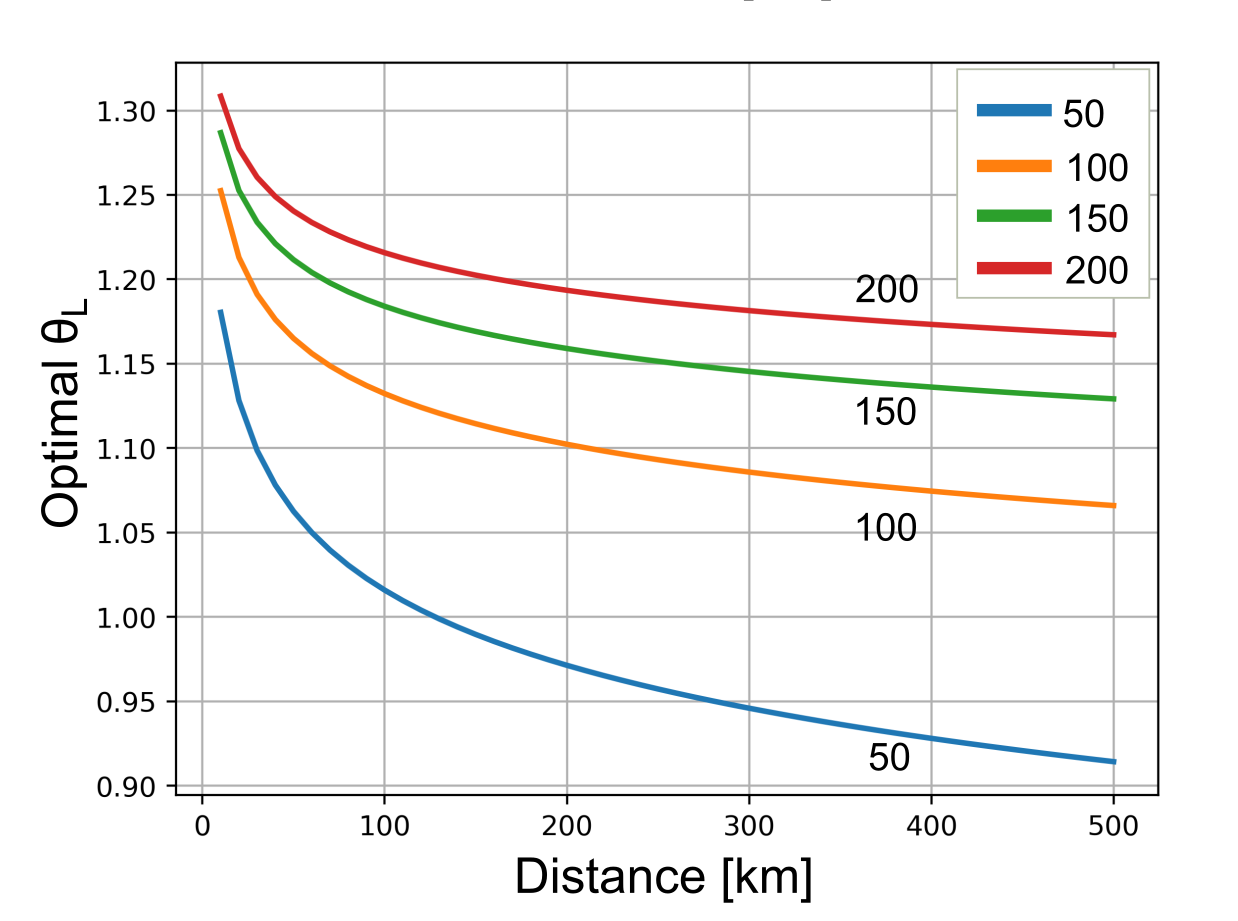}
\caption{\textit{The optimal value of $\theta_L$ for different number of allowed attempts $A$ (legend and plot label) against the distance between the end points of the repeater.}}
\label{Results:Figure:thetaLvsDist}
\end{figure}

\begin{figure}
\centering
\includegraphics[width=1\columnwidth]{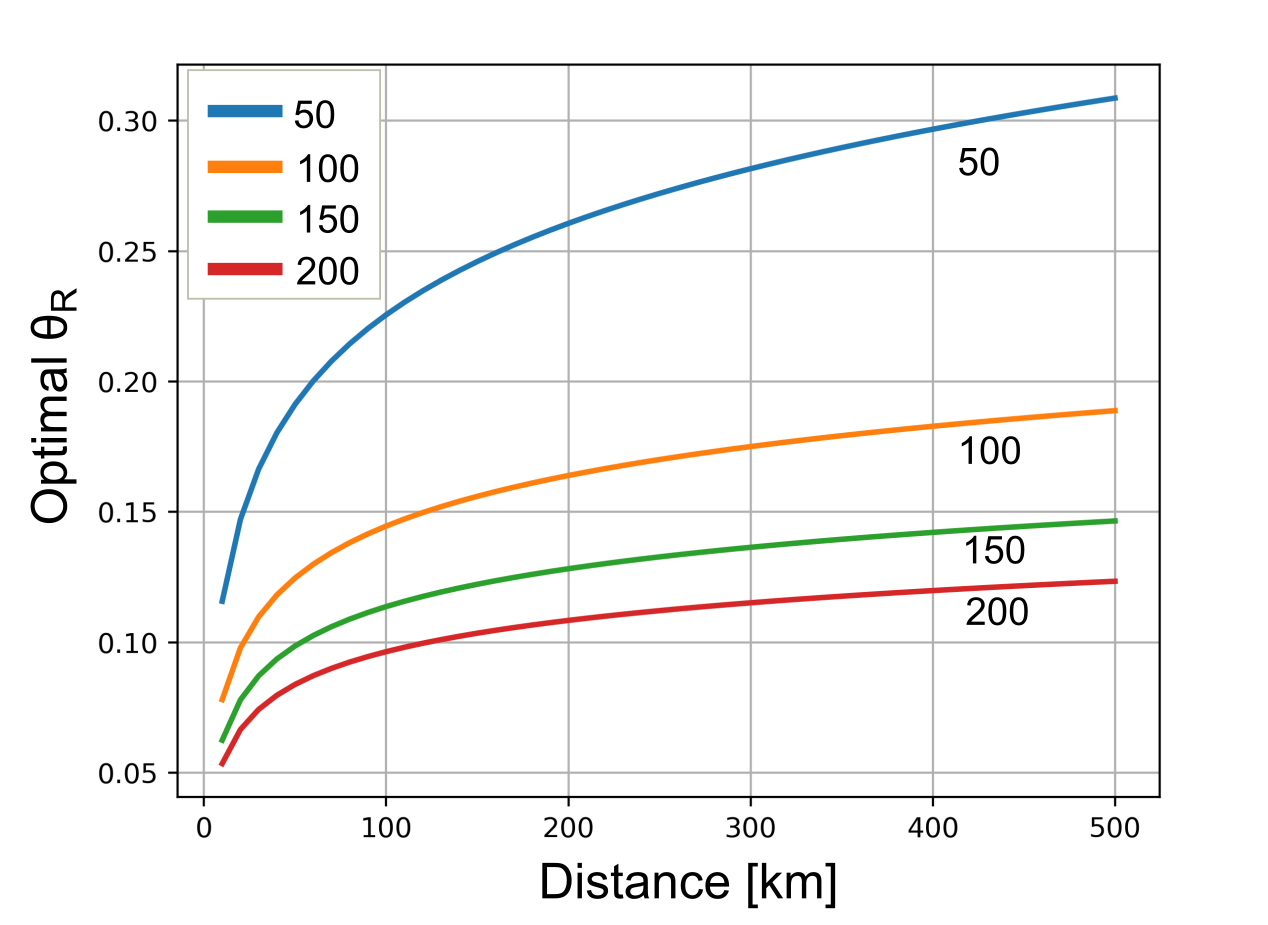}
\caption{\textit{The optimal value of $\theta_R$ for different number of allowed attempts $A$ (legend and plot label) against the distance between the end points of the repeater.}}
\label{Results:Figure:thetaRvsDist}
\end{figure}

In Fig. \ref{Results:Figure:optSep} to the right we plot the critical distance at which the secret key rate vanishes,  against the separation between registers making up a repeater segment. In Fig. \ref{Results:Figure:optSep} to the left we likewise show the critical distance at which the CHSH inequality is no longer broken, also against the separation between registers making up a repeater segment. \\
We then give the optimal average photon number, the optimal values of $\theta_L$, and the optimal values of $\theta_R$ against the distance between the end points of the repeater. These are shown in Fig. \ref{Results:Figure:AvgvsDist}, \ref{Results:Figure:thetaLvsDist}, and \ref{Results:Figure:thetaRvsDist} respectively.
    \label{Analysis:Equation:2qubitState}

\end{document}